\documentclass[journal,10pt,twocolumn]{IEEEtranTCOM}
\ifCLASSINFOpdf
\else
\fi

\usepackage{notation}
\usepackage{graphicx,dblfloatfix}
\usepackage{epsfig}
\usepackage{color}
\usepackage{amsmath}
\usepackage{amssymb}
\usepackage{enumerate}
\usepackage{multirow}
\usepackage{bbm}
\usepackage{floatrow}
\def\G{{\mathcal{G}}}

\newcommand\new[1]{\textcolor{black}{#1}}

\begin{document}

\title{A Scaling Law to Predict the Finite-Length Performance of Spatially-Coupled LDPC Codes}

\markboth{Accepted for publication in IEEE Transactions on Information Theory, ~April, ~2015}{}

\author{Pablo M. Olmos and R\"udiger Urbanke
\thanks{Pablo M. Olmos is with the Dept.~Teor\'i{a} de la Se\~nal
y Comunicaciones,
 Universidad Carlos III de Madrid, Avda. de la Universidad 30,
 28911,
Legan{\'e}s (Madrid), Spain. E-mail: {\tt olmos@tsc.uc3m.es}}
\thanks{R\"udiger Urbanke is with the School of Computer and
Communication Sciences, \' Ecole Polytechnique F\'ed\'erale de
Lausanne, Switzerland.  E-mail: {\tt ruediger.urbanke@epfl.ch}}
\thanks{This work was supported by the European project STAMINA,
265496, by Spanish government MEC TEC2012-38800-C03-01, and by Comunidad de Madrid (project 'CASI-CAM-CM', id. S2013/ICE-2845.
}
\thanks{This paper was presented in part at 2013 IEEE Information Theory Workshop, Sevilla, Spain.}
\thanks{Copyright (c) 2014 IEEE. Personal use of this material
is permitted.  However, permission to use this material for any other purposes must be obtained from the IEEE by sending a request to {\tt pubs-permissions@ieee.org}.
}
}

% make the title area
\maketitle

\begin{abstract} 
Spatially-coupled LDPC codes are known to have excellent asymptotic
properties. Much less is known regarding their finite-length
performance. We propose a scaling law to predict the  error probability
of finite-length  spatially-coupled code ensembles when transmission
takes place over the binary erasure channel. We discuss how the
parameters of the scaling law are connected to fundamental quantities
appearing in the asymptotic analysis of these ensembles and we
verify that the predictions of the scaling law fit well to the data
derived from simulations over a wide range of parameters.  The
ultimate goal of this line of research is to develop analytic tools
for the design of spatially-coupled LDPC codes under practical constraints.
\end{abstract}

% IEEEtran.cls defaults to using nonbold math in the Abstract.
% This preserves the distinction between vectors and scalars. However,
% if the journal you are submitting to favors bold math in the abstract,
% then you can use LaTeX's standard command \boldmath at the very start
% of the abstract to achieve this. Many IEEE journals frown on math
% in the abstract anyway.

% Note that keywords are not normally used for peerreview papers.

% For peer review papers, you can put extra information on the cover
% page as needed:
% \ifCLASSOPTIONpeerreview
% \begin{center} \bfseries EDICS Category: 3-BBND \end{center}
% \fi
%
% For peerreview papers, this IEEEtran command inserts a page break and
% creates the second title. It will be ignored for other modes.
\IEEEpeerreviewmaketitle

\section{Introduction}
\IEEEPARstart{R}{ecently}, it has been proven that spatially-coupled
low-density parity-check (SC-LDPC) codes achieve the capacity of
binary-input memoryless output-symmetric (BMS) channels under
iterative decoding \cite{KudekarBMSIT}.  An SC-LDPC code is constructed
by coupling a chain of $L$ disjoint, or uncoupled, LDPC block codes,
each one of length $M$ bits, together with appropriate boundary
conditions. $L$ is referred to as the SC-LDPC chain length.  Since
spatial coupling is equivalent to introducing memory into the
encoding process, SC-LDPC codes can be viewed as a type of (terminated)
LDPC convolutional code (LDPC-CC) \cite{FelstromZ99}, \cite{Costello14}.

Due to the termination, the  Tanner graph of an SC-LDPC code has a so-called
``structured irregularity'', where parity-check nodes located at
the ends of the chain are connected to a smaller number of variable
nodes than those in the middle \cite{Lentmaier10}.  As a result,
the nodes at the ends of the graph form strong sub-codes and the
resulting reliable information generated there during BP decoding
propagates in a wave-like fashion  from the ends towards the center.
When $M$ tends to infinity and $L$ is sufficiently large, \new{the SC-LDPC code
ensemble exhibits a  BP threshold very close to the maximum-a-posteriori
(MAP) threshold of the uncoupled LDPC code ensemble
\cite{Lentmaier10,KudekarWhy}.} An added feature of SC-LDPC codes
is that  the Tanner graph retains the structure and  properties of the uncoupled  graph. For instance, if the uncoupled LDPC code ensemble 
has a linear growth of the minimum distance
as a function of the block length as is the case for regular LDPC code
ensembles, this property is maintained for the SC-LDPC code ensemble
\cite{Costello14,Mitchell11-2,Sridharan07}.

% The finite-length analysis
%of a particular SC-LDPC ensemble consists of relating the
%BP block error probability to the code parameters, namely the degrees
%$(l,r)$, the chain length $L$, the coupling pattern, and the code
%length $\n=ML$. 

Multiple families of SC-LDPC codes have been proposed and
analyzed to date. In \cite{KudekarBMSIT} and \cite{KudekarWhy},
the authors consider a family of codes generated by coupling together
a series of $(l,r)$-regular LDPC  codes using a random pattern,
determined by the so-called \emph{smoothing parameter} $w$. Coupling 
the codes following a randomized procedure  is convenient for the
asymptotic analysis in terms of  density evolution (DE).  In the limit
$w\rightarrow\infty$,  it was proven that the BP threshold of the
SC-LDPC code ensemble converges to the MAP threshold of the uncoupled
$(l,r)$-regular LDPC   code ensemble.  Alternatively, we can
construct SC-LDPC codes with a certain predefined coupling structure
by means of protographs \cite{Thorpe03}, defining a particular class
of  multi-edge type (MET) ensemble. It has been  shown that
the inherent structure in protograph-based SC-LDPC code ensembles can improve minimum
distance properties \cite{Lentmaier10,Mitchell11-2,Lentmaier10-2}. Also, the
predefined structure allows the use of efficient low-delay windowed
decoding schemes \cite{Iyengar11}.

In  light of all the above, it is probably fair to state that the
asymptotic analysis of SC-LDPC code ensembles is well understood.
Much less is known about their finite-length behavior in the waterfall
region \cite{KudekarWhy}.  Even though there are multiple recent
papers reporting simulation results for different classes of
finite-length SC-LDPC codes
\cite{Tanner04,Hassan12,Pusane08,Bocharova14,Olmos11-3}, to date
we lack analytical models to relate the BP block error probability
of finite-length SC-LDPC codes to their structural parameters.  The
present paper addresses this problem by extending existing analytical
results to analyze finite-length LDPC codes \cite{Urbanke09,Nozaki12}
to a particular family of SC-LDPC codes.

%Namely the check and variable degrees $(l,r)$, the chain length $L$, the coupling pattern, and the code
%length $\n=ML$. 
%Nonetheless, a simulation-based description of the scaling behavior of finite-length SC-LDPC codes over the binary erasure channel (BEC) can be found in 
%\cite{Olmos11-3}.  It is worth to discuss some of the main results. First, if $M$ is kept constant,  the performance degrades with longer chain lengths $L$.  As described in \cite{KudekarBMSIT}, the longer the chain length,
%the more time the ``decoding wave'' needs to sweep over the whole
%chain, since the speed of the decoding wave only depends on the
%channel parameter but not on the length of the chain. Therefore, there is a larger chance that the decoding wave gets ``stuck,'' i.e., that the decoding process fails.
%Beyond describing the scaling behavior using simulations, no closed-form expression was provided
%to evaluate the error probability in the range of practical interest, namely when we consider erasure rates above the BP threshold of the $(l,r)$-regular LDPC code ensemble and below the MAP threshold for the same ensemble. 

We consider a  SC-LDPC code ensemble that is more structured
than the purely random ensemble in  \cite{KudekarWhy} but less structured than the
protograph ensemble.  We denote the ensemble considered here as the
$(l,r,L)$ ensemble. Why do we consider this particular ensemble?
It has been observed empirically that a proper structure improves
performance.  But structure  also makes the analysis harder since
it typically involves more parameters. The chosen ensemble has a
good performance while at the same time is still manageable in terms
of complexity. It is therefore a good compromise. But we  note that the same type of finite-length analysis can
in principle be also performed for other spatially-coupled code ensembles,
and in particular for protograph-based ensembles (using more involved
calculations).  Further, simulation results suggest that the form
of the proposed expression to predict the performance of the $(l,r,L)$
ensemble also applies to protograph-based SC-LDPC code ensembles
if some parameters are adjusted appropriately \cite{Stinner14}.

In \cite{Urbanke09,Amraoui05}, the authors analyze the  finite-length
performance over the binary erasure channel (BEC) of $(l,r)$-regular
LDPC  code ensembles in the waterfall region. For the erasure
channel, the workings of the BP decoder can be cast in an alternative
formulation, namely as peeling decoder (PD) \cite{Luby01}. In this
formulation, any time the value of a variable node is determined,
this variable node and all attached edges are removed from the
graph. In this way we get a sequence of {\em residual} graphs.
Variable nodes are determined either due the received word or later,
during the decoding process, if they are connected to a degree-one
check node.  The analysis of the decoding process consists in
studying the statistical evolution of the residual graph as a
function of time. Indeed, it suffices to analyze the evolution of
the degree distribution (DD) of the residual graph, since this
constitutes a sufficient statistic \cite{Urbanke08-2}.

As shown in \cite{Urbanke09}, the DD of the residual graph at any
time converges (in the code length) to a multivariate Gaussian whose
mean and covariance matrix is given by the solution of a coupled
system of differential equations. Using these results, estimating
the error probability consists in computing the probability that,
during the decoding process, the random process representing the
fraction of degree-one check nodes in the residual graph  reaches
zero before all variables have been determined. For the $(l,r)$-regular
LDPC code ensemble, the error probability is dominated by the
statistics of such a process at a single \emph{critical} point in time.
This critical time is the time at which the expected fraction of
degree-one check nodes in the graph takes on a local minimum
\cite{Urbanke08-2}.

In the present work we extend the described methodology to the
$(l,r,L)$ ensemble.  We derive a system of differential equations
that characterize the statistics of the DD of the residual graph
at any time during the PD process. In particular, the solution of
this system provides the expected number of degree-one check nodes
and the variance around the expected value at any time during the
decoding process.  In contrast to $(l,r)$-regular LDPC code ensembles,
we show that for the $(l,r,L)$ ensemble there exists a \emph{critical
phase} during which both the expected fraction of degree-one check
nodes in the graph and its variance are constant over time.  At any
time during the critical phase, the decoder might fail with uniform
probability. We then proceed in estimating the cumulative error
probability during this critical phase. This requires some effort
since the DD evolution is correlated over time.  More specifically,
we show that during the critical phase the covariance between the
fraction of degree-one check nodes  at two different decoding time
instants  $\tau$ and $\zeta$ decays exponentially with $|\tau-\zeta|$.
Furthermore, the rate of decay is governed by the $(l,r,L)$ coupling
pattern.

Under the assumption that the distribution of the number of degree-one check nodes in the graph is Gaussian
\cite{Urbanke09}, the statistics for  the fraction
of degree-one check nodes during the critical phase of the decoding process are those 
of an appropriately chosen Ornstein-Uhlenbeck (OU) process
\cite{OU-book}.  OU processes have been widely studied and used in
diverse areas of applied mathematics like biological modeling
\cite{OUBio1,OUBio2}, mathematical finance \cite{OUFinance1,OUFinance2}
and statistical physics \cite{OU-3}.  Using known results for the
statistical distribution of the first time at which an OU process
is above certain threshold \cite{OU-1,OU-3,OU-6}, we propose  a
closed-form expression to estimate the error probability of the
SC-LDPC code as a function of $M$, $L$ and the gap to the ensemble's
BP threshold. We illustrate the accuracy of the obtained scaling
law    by comparing with simulated error probability curves.  Further,
the presented closed-form scaling law is consistent with the behavior
we observed by simulations in \cite{Olmos11-3},  capturing the right
scaling behavior between $L$, $M$ and the block error rate.  The
results presented in this paper contribute to a better understanding
of the performance of finite-length SC-LDPC codes and  the  proposed
scaling law constitutes a useful engineering tool for code design.
In particular, it can be used to accurately estimate the performance
improvement/degradation when a certain parameter of the SC-LDPC code
is modified.

The paper is structured as follows. In Section\SEC{S1} we review
the construction of the SC-LDPC code ensemble that is analyzed in
the rest of the paper. In Section\SEC{PD}, we derive the differential
equations that describe the expected graph evolution during PD.
Graph covariance evolution is discussed in Section\SEC{COVPD}. In
Section\SEC{SL}, we present the scaling law to predict the SC-LDPC
block error rate and, finally, in Section\SEC{Conclusions}
we provide some concluding remarks and potential future lines of
research.

\section{Construction of the SC-LDPC code ensemble}\LABSEC{S1}

Consider a set of $L$  uncoupled  $(l,r)$-regular  LDPC block codes
of length $M$ bits, where $l$ is the variable-node degree and $r
\geq l$ is the check-node degree.  The Tanner graph of each individual
code has $M$ variable nodes, each of degree $l$, and $\frac{l}{r}M$
(assume it to be integer) check nodes, each of degree $r$, and the
(design) code rate of each  code is $1-\frac{l}{r}$.  From this set
of codes we construct an SC-LDPC code by permuting edges in such a
way that the individual codes become connected (coupled) but maintain
their original degree distribution.  Let us describe this procedure
in more detail.

The SC-LDPC Tanner graph has one position for each of the $L$
uncoupled codes, index these positions by $u$, $u=1,\ldots,L$.
Hereby, position $u$ contains the set of $M$ variable nodes and
$\frac{l}{r}M$ check nodes that originally belonged to the $u$-th
uncoupled code.

We consider a very simple and regular coupling pattern that is
convenient for the purpose of analysis, and we refer to the
corresponding ensemble as the $(l,r,L)$ ensemble. A variable node
at position $u$ has exactly one connection to a check node at
position $u+i$, $i=0, \cdots, l-1$.  For instance, for the case
$l=3$, a variable node at position $5$ is connected to one check
node at position $5$, one check node at position $6$ and one check
node at position $7$. To maintain this pattern for all bits in the
code,  note that $l-1$ extra positions, containing exclusively check
nodes, are needed at the end of the chain. Consequently, the $(l,r,L)$
graph has $D=L+l-1$ positions with $\frac{l}{r}M$ check nodes each.
% not clear if we need this Assume all check nodes in the graph
Each check node in the graph has maximum degree $r$ and hence there are $r\times\frac{l}{r}M=lM$
check node sockets per position.

\new{ To generate a code from the $(l,r,L)$ ensemble,  we proceed
as follows. First, we label the set of check node sockets per
position from $1$ to $lM$ and the set of variable nodes per position
from $1$ to $M$.  Then, for $u=1,\ldots,D$ 
\begin{enumerate} 
\item
Perform a random permutation of the set $\{1,\ldots,lM\}$,
$\pi_u=\text{randperm}\{1,\ldots,lM\}$. The permutation is selected
with uniform probability among all possible $lM!$ permutations.
\item Divide $\pi_u$ into $l$ disjoint sets of $M$ elements each.
They are denoted by $\pi_u^{0},\pi_u^{1},\ldots,\pi_u^{l-1}$. 
 \item
For $i=0,\ldots,l-1$, we place $M$ edges in the graph connecting
the $M$ variable nodes  at position $u-i$  with the $M$ check node
sockets in position $u$ with labels contained\footnote{In other
words, for $m=1,\ldots,M$, we place an edge in the graph that
connects the $m$-th variable node in position $u-i$ to the check
node socket in position $u$ whose label corresponds to the $m$-th
element in the set $\pi_u^{i}$.} in the set $\pi_u^{i}$.   If $u-i<0$
or $u-i>L$, the check node sockets in position $u$ with labels
contained in $\pi_u^{i}$ are left empty.  
\end{enumerate} }

From the perspective of the variable nodes, the ensemble looks like
the protograph-based  ensemble  considered in \cite{Lentmaier10}.
However, from the perspective of the check nodes, the ensemble is
closer to the random ensemble considered in \cite{KudekarWhy}. As
mentioned, we chose this ensemble mainly since it is easy to analyze.

\subsection{Ensemble properties and design rate }
Any code in the $(l,r,L)$ ensemble has the following properties:
\begin{enumerate}
\item[I)] Every variable node at position $u$, $1\leq u\leq L$, is
connected via a single edge to a check node at position $u+i$,
for $i=0,\ldots,l-1$.
\end{enumerate}
\begin{enumerate}
\item[II)] If we select at random an edge that is connected to a check node at position $u$, $l\leq u \leq L$, then the
position of the variable node  connected to this edge is a random
variable uniformly distributed in the interval $\{u-(l-1),\ldots,u\}$.
For $1\leq u \leq l-1$ and $L+1\leq u \leq D $, the same statement
holds but in the intervals $\{1,\ldots,u\}$ and $\{u-(l-1),\ldots,L\}$
respectively.
\end{enumerate}
\begin{enumerate}
\item[III)] All check nodes at position $u$, $l\leq u \leq L$, have degree $r$.
\end{enumerate}
\begin{enumerate}
\item [IV)] The degree  of a check node picked at random at position
$u$, $1\leq u \leq l-1$, is a random variable distributed according
to a Binomial distribution of $r$ trials and probability of success
$u/l$. Similarly, the degree of a check node picked at random at
position $u$, $L+1\leq u \leq D$, is a random variable distributed
according to a Binomial distribution of $r$ trials and probability
of success $(L+l-u)/l$.
\end{enumerate}

Properties I), II) and II) are easy to check. Property IV) is a
direct consequence of the step 3) in the construction of the code
described above. For $1 \leq u \leq l-1$, check node sockets in
position $u$ with labels in the sets
$\pi_u^{0},\pi_j^{1},\ldots,\pi_j^{u-1}$ are occupied while check
node sockets with labels in the sets
$\pi_u^{u},\pi_j^{u+1},\ldots,\pi_j^{l}$  are left empty. This means
that, among the $lM$ check node sockets,  $(l-u)M$ of them (chosen
at random) are left empty. Thus, every check node socket in this
position is left empty with probability $(l-u)/l$. Accordingly, the
degree of  a check node chosen at random in this position is given by a Binomial distribution
with  $r$ trials and  success probability $u/l$.

To compute the design rate of the $(l,r,L)$ ensemble, we need to find the average
number of check nodes in the graph connected to at least one variable
node.  As discussed above, each one of the $\frac{l}{r}M$ check
nodes at positions $l,\ldots,L$ are of degree  $r$. By property
IV), a check node at position $u$, $1\leq u \leq (l-1)$, is of
degree at least one with probability

\begin{align}
1-\left(\frac{l-u}{l}\right)^r
\end{align}
and thus the average number of check nodes with degree at least one
in the graph is given by
$\frac{l}{r}M\left(L-(l-1)+2\sum_{u=1}^{l-1}1-\left(\frac{l-u}{l}\right)^r\right)$.
Therefore, the design rate of the ensemble is
\begin{align}
\rate&=1-\frac{\frac{l}{r}M\left( L-(l-1)+2\displaystyle\sum_{u=1}^{l-1}1-\left(\frac{l-u}{l}\right)^r\right)}{ML}\nonumber\\\nonumber\\\LABEQ{RATELDPCC}
%&=1-\frac{l}{r}\frac{\displaystyle L-l+2\sum_{j=1}^{l-1}1-\left(\frac{j}{l}\right)^r}{L}\\
&=1-\frac{l}{r}\left(1-\frac{l-1}{L}+2\frac{\displaystyle\sum_{u=1}^{l-1}1-\left(\frac{l-u}{l}\right)^r}{L}\right)
\end{align}
This tends to $1-l/r$, the rate of the uncoupled $(l,r)$-regular
LDPC  code ensemble, when $L\rightarrow\infty$.

\section{Peeling Decoding and Expected Graph Evolution}\LABSEC{PD}
Consider transmission over the BEC and decoding using the peeling
decoder \cite{Urbanke08-2}. At each step, the PD removes one
degree-one check node and its connected variable node, as well as
all edges connected to these two nodes. As a result, the PD 
process gives rise to a  sequence of residual graphs. In
\cite{Luby01},  it is shown that if we apply the PD to elements of
an LDPC code ensemble, the expected sequence of graphs or \emph{expected
graph evolution}  can be computed  by solving a system of coupled
differential equations. Note that the expectation is done here with respect
to the channel realization as well as with respect to the code
ensemble. Note further that standard arguments show that most codes
and most channel realizations lead to a decoding behavior that is
close to this expected behavior \cite{Urbanke08-2}. Therefore, the solution of the
differential equation encodes the asymptotic performance, in
particular the threshold of the LDPC code ensemble.

In \cite{Urbanke09}, \emph{scaling laws} (SLs) were proposed to
predict the BP finite-length performance for the $(l,r)$-regular LDPC code ensemble in the waterfall
region.\footnote{The above analysis only captures ``large'' decoding
failures, i.e., decoding failures where a linear fraction of the
bits remains undecoded. Such decoding failures are dominant for
channel parameters close to the threshold of the code and lead to
the characteristic ``waterfall''-shape of the error probability
curve in this regime.  In addition, the decoder can ``essentially''
succeed but might fail to decode a few remaining bits.  This is the
dominant failure mode for channel parameters sufficiently away from the
threshold and the mechanism for this failure is of an entirely
different nature.  The resulting shape of the error probability curve
in this regime is called the ``error floor.''} It was also shown
how to compute the parameters of the scaling law from the DD of the
ensemble.  The four most important scaling parameters are (i) the
threshold, (ii) the ``critical time(s)'' of the process, i.e., the
time at which the expected number of degree-one check nodes reaches
a local minimum, (iii) the expected number of degree-one check nodes
at the critical time(s), and (iv) the variance of this quantity at
the critical time(s). The first three parameters can be determined
by looking at the expected graph evolution.  To determine the fourth
parameter, a further system of differential equations, dubbed
covariance evolution, has to be solved  \cite{Urbanke09}.  For
$(l,r)$-regular LDPC code ensembles, there is only a single critical
time and the error probability can be estimated from the relationship
between the expected number of degree-one check nodes  and the
variance of the same quantity at this point in time.

In the following, we apply a similar line of reasoning to the
$(l,r,L)$ ensemble.  In Section\SEC{PDEXPECTED}, we compute the
expected graph evolution of the $(l, r L)$ ensemble and we discuss
the first three scaling parameters (i)-(iii). In Section\SEC{COVPD},
we derive the system of differential equations of the graph covariance
evolution.

%\begin{figure*}[!b]
%\centering 
%\begin{tabular}{ccc}
%\includegraphics[scale=0.32]{fig2_var2.eps} & \includegraphics[scale=0.32]{fig2_var1.eps}  &
%\includegraphics[scale=0.32]{fig2_corr.eps}\\
%(a) & (b) & (c)
%\end{tabular}
%\caption{In (a), we plot the empirical estimate to $M\delta_1(\tau)$
%obtained after averaging $10^4$ samples for $\epsilon=0.45$ $(\lhd)$
%and $\epsilon=0.46$ $(\circ)$. We consider two $M$ values: $M=1400$
%(dashed lines) and $M=2600$ (solid lines). In (b), we normalize
%these curves by $\Delta_{\pe}$. In (a), we also include the case $L=100$
%$M=1400$ and $\pe=0.45$, green dashed lines with $*$ marker.
%In (c), for the same data set we plot the empirical $M
%\phi_{1}(\tau,\tau+\zeta)$ for $\zeta=15$. }\LABFIG{FigVAR}
%\end{figure*}

\subsection{Description of the $(l,r,L)$ ensemble in terms of degree distribution}\LABSEC{PDEXPECTED}
Our aim is to describe the statistical graph evolution of elements
in the $(l,r,L)$ ensemble during the decoding process. In the sequel
we denote time by $\ell\in\mathbb{N}$ if we measure time in discrete
steps and by $\tau$ once we rescale time to pass to the continuous
limit in order to write down a differential equation, where
$\tau\doteq\ell/M$.  Since the PD removes one variable-node per
iteration, in average -- the erasures are random -- we need $\pe L M$ iterations to success and
thus $\tau\in[0,\epsilon L]$.

At time $\ell$, let $R_{j,u}(\ell)$ be the number of edges that are
connected to check nodes of degree $j$, $j=1,\ldots,r$, placed at position $u$, $u=1,\ldots,D$, and
let $E_u(\ell)$ be the total number of edges connected to check
nodes at this position.  Thus,
\begin{align}
E_u(\ell)=\sum_{j=1}^{r}R_{j,u}(\ell).
\end{align}
For the $(l,r,L)$ ensemble all variables nodes are of
degree $l$ and  it suffices therefore to know the number of remaining
variable nodes per position at iteration $\ell$. Let this number
be denoted as $V_u(\ell)$. Note also that the following relation
holds
\begin{align}
l\sum_{u=1}^{L}V_u(\ell)=\sum_{u=1}^{D}E_u(\ell)=\sum_{u=1}^{D}\sum_{j=1}^{r}R_{j,u}(\ell)\doteq E(\ell).
\end{align}

\subsection{$(l,r,L)$ expected graph evolution}\LABSEC{PDEXPECTED}
Assume that we use samples from the $(l,r,L)$ ensemble to transmit over
a BEC with erasure probability $\pe$.  We initialize the PD by
removing from the graph all variable nodes whose value was received
through the channel as well as all their edges. 
% At each iteration, the PD then choses a degree-one check node. This check
%node is removed along with the connected variable node and all
%connected edges.
%
Let $\ell=0$ be the state of the system after this initialization.
%and let $V_u(\ell)$ and $R_{j,u}(\ell)$, $j=1,\ldots,r$ and
%$u=1,\ldots,D$, be the DD at position $u$ after $\ell$ PD iterations.
At $\ell=0$, the expected number of variables per position of the graph is
\begin{align}\LABEQ{INIT1}
\E[V_u(0)]&=\left\{\begin{array}{cc} \pe M, & 1\leq u \leq L, \\ 0, & \text{otherwise},\end{array}\right .
\end{align}
and the expected value of $R_{j,u}(0)$ is given by
\begin{align}\LABEQ{INIT2}
%\E[R_{j,u}(0)]&=j\sum_{j}\binom{r}{j}(\pe \frac{d_u}{l})^{j}(1-\pe \frac{d_u}{l})^{(r-j)},
\E[R_{j,u}(0)]&=j\frac{l}{r}M\sum_{m\geq j}^{r} \rho_{m,u} \binom{m}{j} \pe^j (1-\pe)^{(m-j)},
\end{align} 
where $\rho_{m,u}$ is the probability that a check node chosen at
random from position $u$ in the original $(l,r,L)$ code graph is
of degree $m$. The term that multiplies $\rho_{m,u}$ inside the sum
in \EQ{INIT2} corresponds to  the probability that $m-j$ edges are removed from a degree-$m$ check node after PD initialization, hence it becomes a degree-$j$ check node.

%
%that a degree-$m$
%check node ``looses'' $m-j$ edges during the PD initialization and hence
%becomes a degree-$j$ check node.

According to the description of the $(l,r,L)$ ensemble and the list
of properties described in Section\SEC{S1}, all check nodes at
intermediate positions are of degree $r$ and thus, for $l\leq u
\leq L$, $\rho_{r,u}=1$ and $\rho_{m,u}=0$, $m<r$. For $u=1,\ldots,l-1$,
$\rho_{m,u}$ is a Binomial p.m.f. of the form
\begin{align}
\rho_{m,u}=\binom{r}{m}  \left(\frac{u}{l}\right)^m\left(1- \frac{u}{l}\right)^{(r-m)},
\end{align}
and at the opposite boundary the same quantities can be found by symmetry. Namely, for $u=1,\ldots,l-1$
\begin{align}
\rho_{m,L+l-u}=\rho_{m,u}.
\end{align}

Define the normalized DD at time $\tau$ as follows
\begin{align}\LABEQ{norm}
 r_{j,u}(\tau)\doteq\frac{R_{j,u}(\ell)}{M} \quad \text{ and } \quad v_{u}(\tau)\doteq\frac{V_u(\ell)}{M}
\end{align}
and let $\mathcal{G}(\tau)=\{\mathcal{G}_{j,u}(\tau)\}$ for $u=1,\ldots,D$ and $j=1,\ldots,r+1$ be a compact notation for the set of all DD coefficients, where 
\begin{align}
\mathcal{G}_{j,u}(\tau)&=\left\{
\begin{array}{cc}
r_{j,u}(\tau) & j\in\{1,\ldots,r\}\\
v_u(\tau) & j=r+1
\end{array}
\right. .
\end{align}
Also, to keep the notation uncluttered, in the following we write
$[r+1]$ to denote $\{1,\ldots,r+1\}$ and $[D]$ to denote $\{1,\ldots,D\}$.

The expected graph evolution of the $(l,r,L)$ ensemble is determined
by $\E[\mathcal{G}(\tau)]$ for $\tau\in[0,\epsilon L]$.
%, since the DD is a sufficient statistic for the sequence of residual graphs. 
As shown in \cite{Luby01,Urbanke08-2}, the  system of differential equations
\begin{align}\LABEQ{system1}
\frac{\partial \hat{\mathcal{G}}_{j,u}(\tau)}{\partial \tau}&=f_{j,u}\left(\hat{\mathcal{G}}(\tau)\right)
\end{align}
for $u\in[D]$ and $j\in[r+1]$, where
\begin{align}\LABEQ{system12}
f_{j,u}\left(\mathcal{G}(\tau)\right)&\doteq\frac{\E\left[\mathcal{G}_{j,u}(\tau+\frac{1}{M})-\mathcal{G}_{j,u}(\tau)\Big| \mathcal{G}(\tau) \right] }{1/M},
\end{align}
has a unique solution and, further, the solution for the initial
conditions $\hat{\G}_{r+1,u}(0)=\E[V_u(0)]/M$ and
$\hat{\G}_{j,u}(0)=\E[R_{j,u}(0)]/M$, given in
\EQ{INIT1} and \EQ{INIT2}, deviates from the true mean  evolution
of $\G(\tau)$ by less than $M^{-1}$. In addition, with probability
$1-\mathcal{O}(\text{e}^{-\sqrt{M}})$, any sample of $\G$ deviates
from $\hat{\G}$ by less\footnote{This concentration result is based
on the analysis of the evolution of (martingale) Markov processes
due to Wormald \cite{Wormald95}. In the Wormald method we do not
look at a single Markov process but a sequence of such processes
parameterized by some quantity $m$, $\{Z^{(m)}(t)\}_{m\geq 1}$.
The Wormald method consists of showing that, for increasing $m$,
with high probability the random variables $Z^{(m)}(t=0)$,
$Z^{(m)}(t=1)$, \ldots ~ stay close to an \emph{expected evolution} and
that this evolution is equal to a solution of a differential equation.
In our context, the Markov process of interest is the DD at any
time $\ell$ and we consider sequences of processes corresponding
to the residual DD of the $(l,r,L)$ ensemble with increasing $M$
values, which motivates using $M$ to normalize the DD in \EQ{norm}
following Wormald's method. Any quantity proportional to $M$
guarantees concentration around the expected graph evolution and
thus could be chosen to normalize the DD.  For instance, the total
code length $\n=ML$ or the expected number of bits per position
after PD initialization $\epsilon M$.} than $M^{-1/6}$ \cite{Urbanke08-2}.
Consequently,  in the limit $M\rightarrow\infty$, samples of the
process $\G(\tau)$ closely follow  $\hat{\G}(\tau)$.  Note that the
function $f_{j,u}\left(\mathcal{G}(\tau)\right)$ represents the
``drift'' (expected change) in the components of the DD as a result
of one decoding step.  The computation of the expectations in
\EQ{system12} for the $(l,r,L)$ ensemble can be found in
Appendix\SEC{A1}.

To evaluate the  probability of successful
decoding, we have to estimate the probability that the random 
process representing the total fraction of degree-one check nodes in the graph, namely
\begin{align}\LABEQ{r1}
r_{1}(\tau)\doteq\sum_{u=1}^{D}\G_{1,u}(\tau)=\sum_{u=1}^{D}r_{1,u}(\tau),
\end{align}
stays strictly positive until the whole graph has been peeled
off. Note that the ensemble BP threshold $\pe_{(l,r,L)}$ is defined
as the maximum $\pe$ for which the expected fraction of degree-one
check nodes in the graph, i.e.,
\begin{align}\LABEQ{hatr1}
\hat{r}_{1}(\tau)\doteq\sum_{u=1}^{D}\hat{\G}_{1,u}(\tau)=\sum_{u=1}^{D}\hat{r}_{1,u}(\tau),
\end{align}
is strictly positive for any $\tau\in[0,\epsilon L]$. 

\begin{figure}[h]
\centering \includegraphics[scale=0.5]{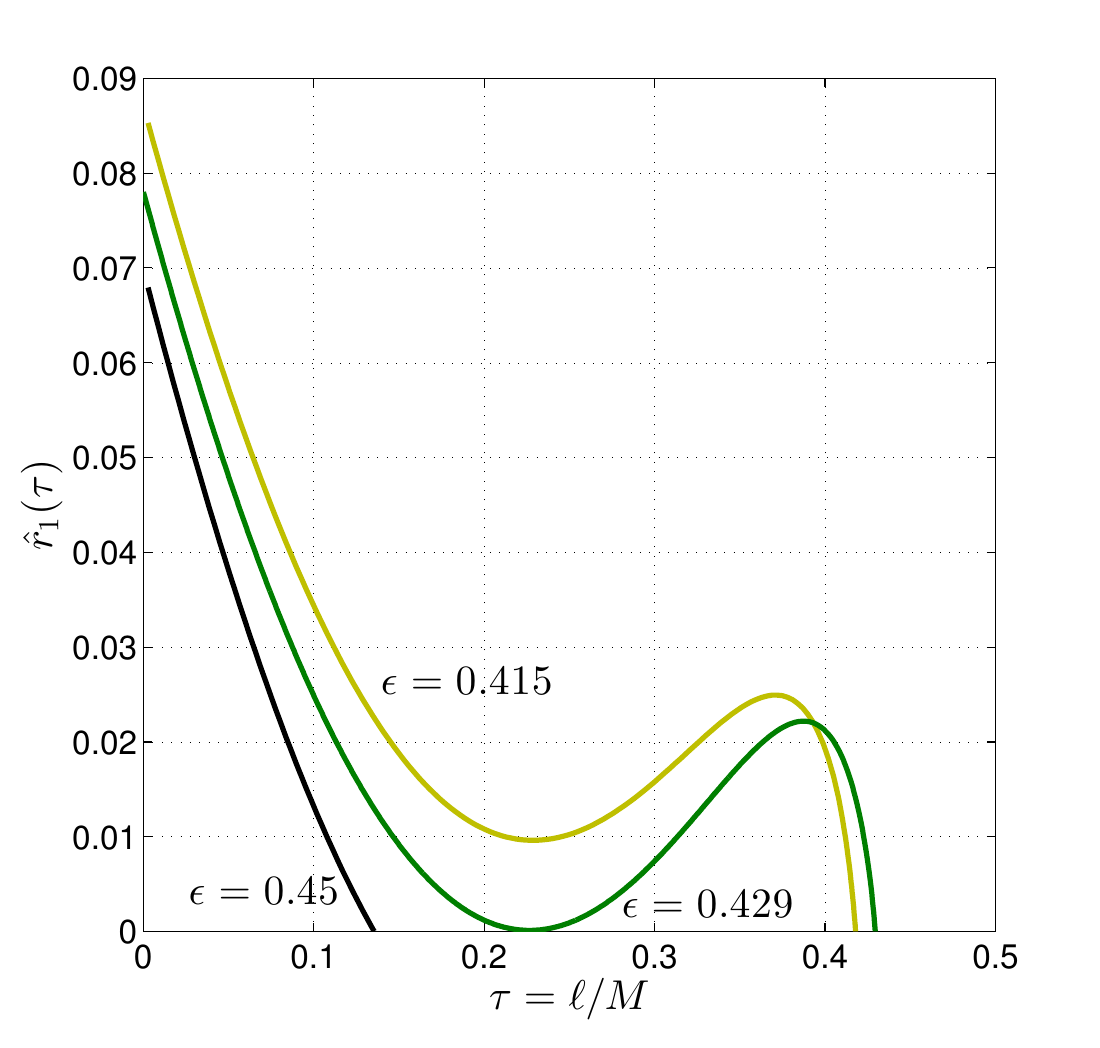}
\caption{Evolution of the expected fraction of degree-one check nodes in the residual graph as the PD iterates for the $(3,6)$-regular LDPC code ensemble at $\epsilon=0.415$, $\epsilon=0.429$ and $\epsilon=0.45$. }\LABFIG{36evol} 
\end{figure}

\begin{figure*}[!b]
\centering
\begin{tabular}{cc}
\includegraphics[scale=0.45]{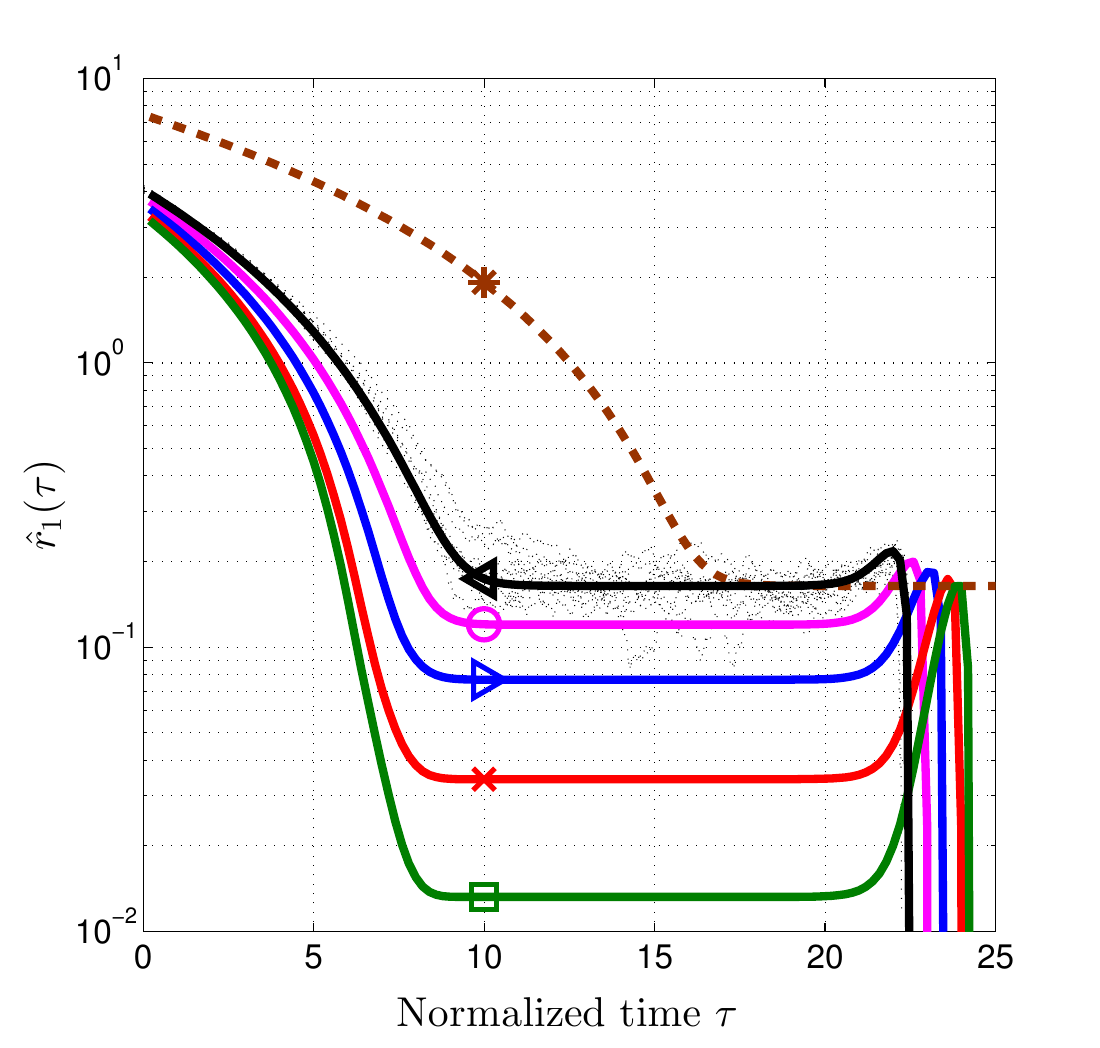}  & \includegraphics[scale=0.45]{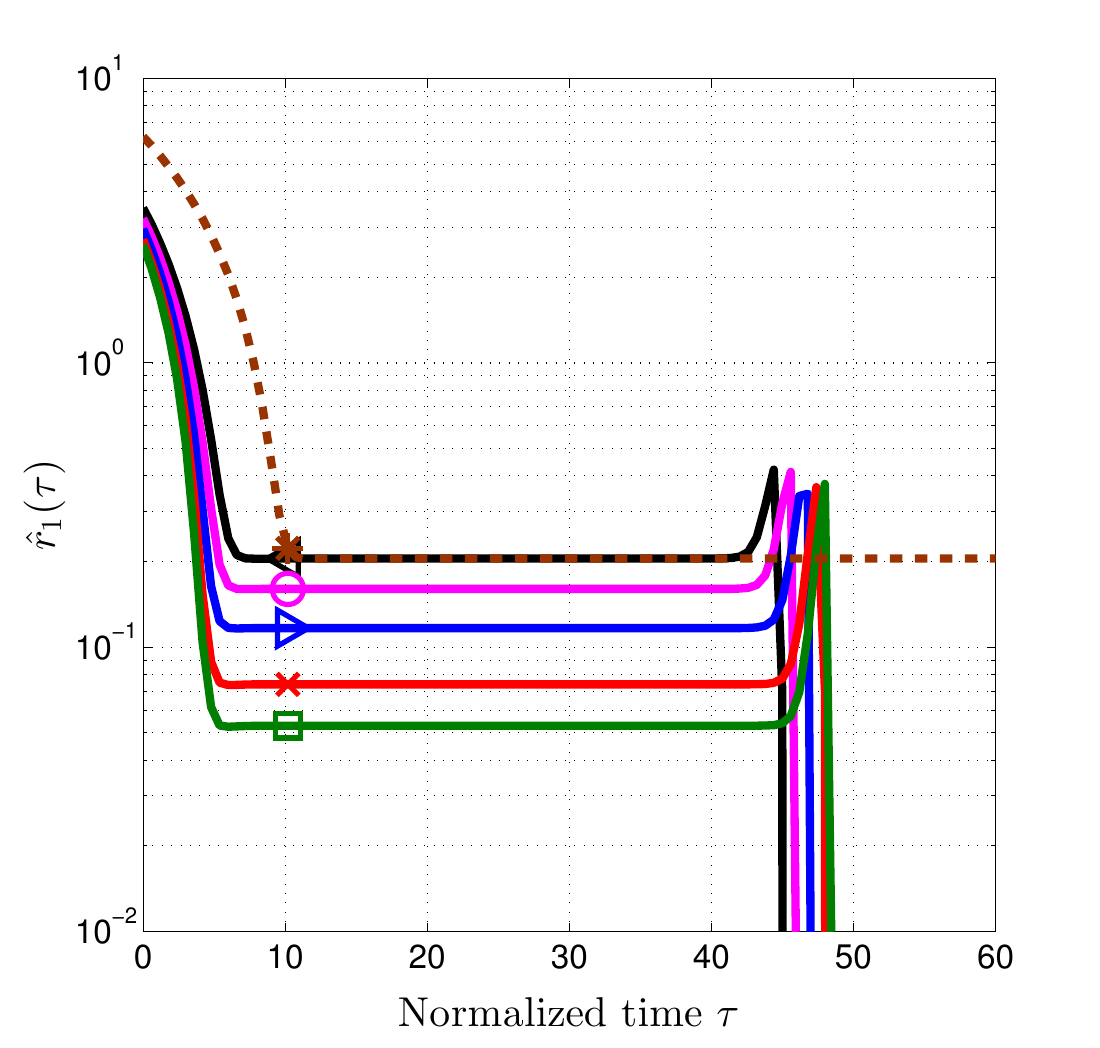}  \\
(a) & (b)
%\\
%\includegraphics[scale=0.5]{fig2_unlinked.eps}  & \includegraphics[scale=0.5]{fig482_unlinked.eps}  \\
%(c) & (d)
\end{tabular}
\caption{ $\hat{r}_{1}(\tau)$ in \EQ{hatr1} for the $(3,6,50)$ ensemble
(a) and for the $(4,8,100)$ ensemble (b). In both figures, the $\pe$
values considered are:  $0.45$ $(\lhd)$, $0.46$ $(\circ)$, $0.47$
$(\rhd)$, $0.48$ $(\times)$ and $0.485$ $(\square)$. In (a), for
$\pe=0.45$ we have included a set of $10$ empirical trajectories
computed for $M=1000$ bits. They are shown as dashed thin lines. For the same ensembles
and $\epsilon=0.45$, we also include $\hat{r}_{1}(\tau)$ for the
double chain length, namely $L=100$ and $L=200$ respectively, dashed
lines with $(\ast)$ marker.
%In (b), we normalize the $\hat{r}_{1}(\tau)$ curves by
%$\Delta_{\pe}$. The thresholds are respectively $\pe_{3,6,50}=0.48815$ and $\pe_{4,8,100}=0.49774$. 
 } \LABFIG{FigDE} \end{figure*}

\subsection{Solution and comparison for different ensembles}\label{S1}

Before discussing the solution to $\hat{r}_1(\tau)$ in \EQ{hatr1}
for the $(l,r,L)$ ensemble, it is worth showing the corresponding
solution for the uncoupled $(l,r)$-regular LDPC code ensemble. In
Fig. \FIG{36evol}, we represent the evolution of the expected
fraction of degree-one check nodes $\hat{r}_{1}(\tau)$  for a
$(3,6)$-regular LDPC code ensemble  at $\epsilon=0.415$, $\epsilon=0.429$
and $\epsilon=0.45$. For this code ensemble, $\hat{r}_{1}(\tau)$
as a function of $\tau$ is  known in closed-form \cite{Luby01}.
The BP threshold of the $(3,6)$-regular LDPC code ensemble is
$\pe_{(3,6)}=0.4294$. Observe that the expected evolution has a
single local minima or \emph{critical point}. Indeed, the threshold
is that $\pe$ parameter where at the critical point the curve is tangent to the x-axis. Above $\pe_{(3,6)}=0.4294$, $\hat{r}_{1}(\tau)$
becomes zero before the whole graph has been peeled off. Therefore,
at erasure rates above the threshold, with high probability the PD
over any element of the $(3,6)$-regular LDPC code ensemble will not
succeed \cite{Urbanke09}.

Let us now discuss the spatially-coupled ensemble.  In Fig.~\FIG{FigDE},
we plot the solution for $\hat{r}_{1}(\tau)$ for the $(3,6,50)$
ensemble (a) and for the $(4,8,100)$ ensemble (b) for different
$\pe$ values. The quantity $\hat{r}_{1}(\tau)$ is computed via
numerical integration of the differential equations in \EQ{system1}
using Euler's method. For the same ensembles and $\epsilon=0.45$,
we also include $\hat{r}_{1}(\tau)$ when the chain length is doubled,
i.e., when $L=100$ in (a) and $L=200$ in (b). In Fig.  \FIG{FigDE}(a),
for the case $\pe=0.45$, we have further included a set of $10$
simulated decoding trajectories computed for $M=1000$ bits to show
that they indeed concentrate around the predicted  evolution.

The evolution of $\hat{r}_{1}(\tau)$ shown in Fig.~\FIG{FigDE}
shows three distinct stages that we now briefly discuss:

\subsubsection{Initial phase}
Similar to what we have seen in Fig.~\FIG{36evol}, we can observe
in Fig.~\FIG{FigDE}(a) and (b) an initial phase of a rapid decay
in degree-one check nodes. This phase starts right after the
initialization and corresponds to a phase where there are many
degree-one check nodes more or less uniformly spread out across the
length of the chain. During this phase the bulk of the system behaves
essentially like the uncoupled system  and only at the boundaries
do we see small deviations from this behavior due to the termination.
Consequently, degree-one check nodes are removed roughly uniformly
along the length of the chain. This can be observed by plotting the
average probability $p_u(\tau)$ that the PD removes a degree-one
check node from position $u$ at time $\tau$:
\begin{align}\LABEQ{pu}
p_u(\tau)\doteq\frac{\hat{r}_{1,u}(\tau)}{\displaystyle \sum_{m=1}^{D}\hat{r}_{1,m}(\tau)}=\frac{\hat{r}_{1,u}(\tau)}{\hat{r}_{1}(\tau)}
\end{align}
for $u\in[D]$. In Fig.~\FIG{ProfilePU}, we plot in solid lines the
$p_u(\tau)$  profile for the $(3,6,50)$ ensemble for $\pe=0.45$ at
two time instants: $\tau=5$ ($\triangledown$) and $\tau=15$
$(\diamond)$. The dashed lines represent the profile of variable
nodes $\hat{v}_u(\tau)$ per position, $u=1,\ldots,L$, at the same
two instants, $\tau=5$ ($\square$) and $\tau=15$ ($\circ$). At
$\tau=5$, the decoder is in the initial phase and we can observe that
at this point in time $p_u(\tau)$ is approximately uniform (with a
small extra bump at the boundaries due to the termination).  Indeed,
based on $p_u(\tau)$, we can compute that the cumulative probability
of removing a degree-one check node from positions $1-4$ or $48-52$
is less than 0.3.

\subsubsection{Second phase: wave-like decoding} If we are transmitting
at a channel value which is strictly above the BP threshold of the
uncoupled $(l,r)$-regular LDPC code ensemble, then the initial phase
ends when all positions except those at the boundaries have run
out of degree-one check nodes. At this time, denoted by 
$\tau^*=\tau^*(l,r,L,\epsilon)$, the ``interior''  of the coupled
system (i.e., the positions away from the boundaries) has reached a defacto fixed
point and this fixed point is equal to the fixed point that the
uncoupled system reaches with the same channel parameter
\cite{KudekarWhy}.  Only towards the two boundaries are there still
some degree-one check nodes available and those keep the decoding
``alive''. In Section\SEC{lower}, we show how to compute a lower-bound
on $\tau^*$ based on the expected graph evolution for the uncoupled
$(l,r)$-LDPC code ensemble.

\begin{figure}[h]
\centering \includegraphics[scale=0.45]{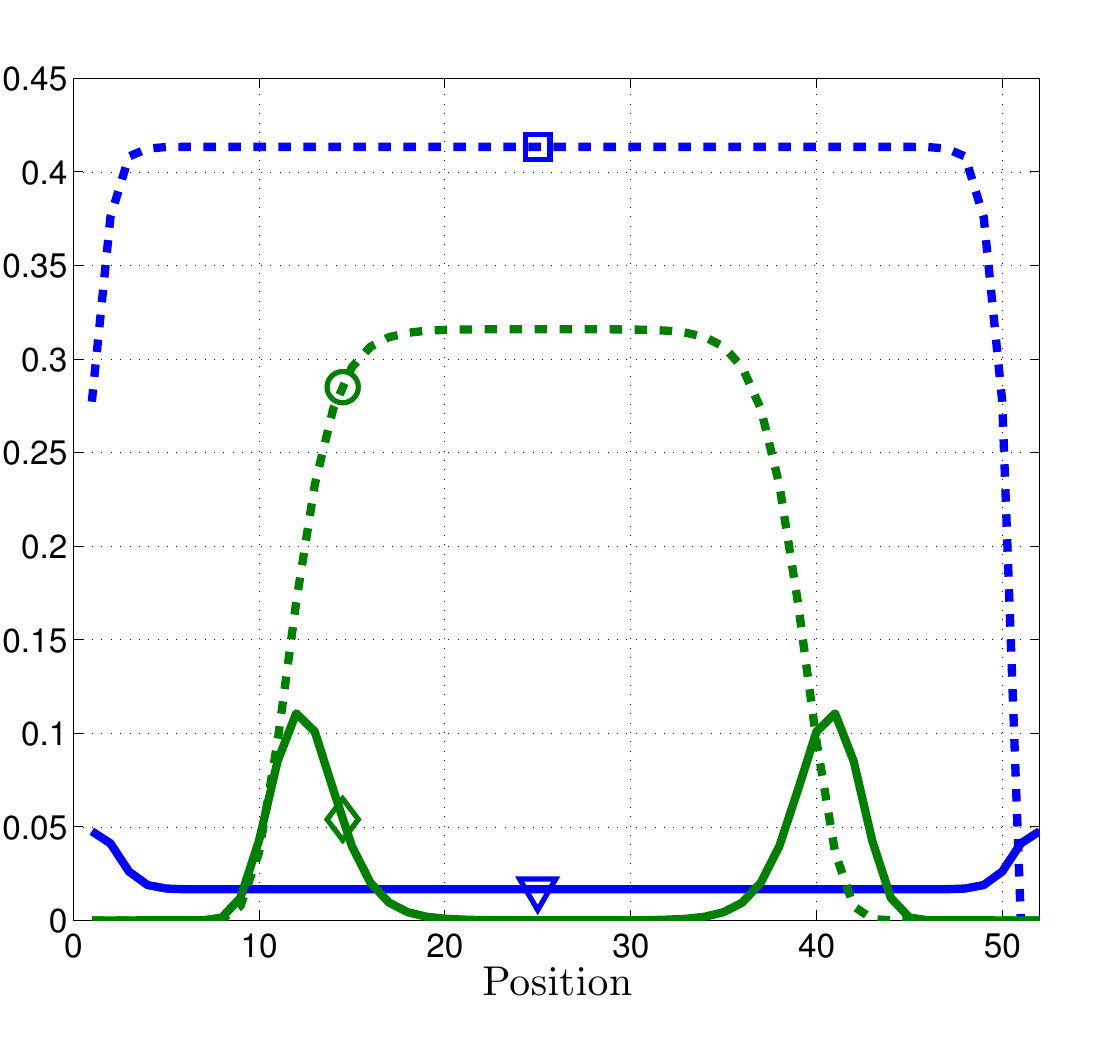}
\caption{ For the $(3,6,50)$ ensemble,
we plot in solid lines the $p_u(\tau)$ profile at $\tau=5$ ($\triangledown$) and $\tau=15$ $(\diamond)$ for $\pe=0.45$. Dashed lines represent the expected profile of variable nodes per position $\hat{v}_{u}(\tau)$ at the same time instants: $\tau=5$ ($\square$) and $\tau=15$ ($\circ$). }\LABFIG{ProfilePU} \end{figure}

\begin{figure*}[!b]
\centering
\begin{tabular}{cc}
\includegraphics[scale=0.45]{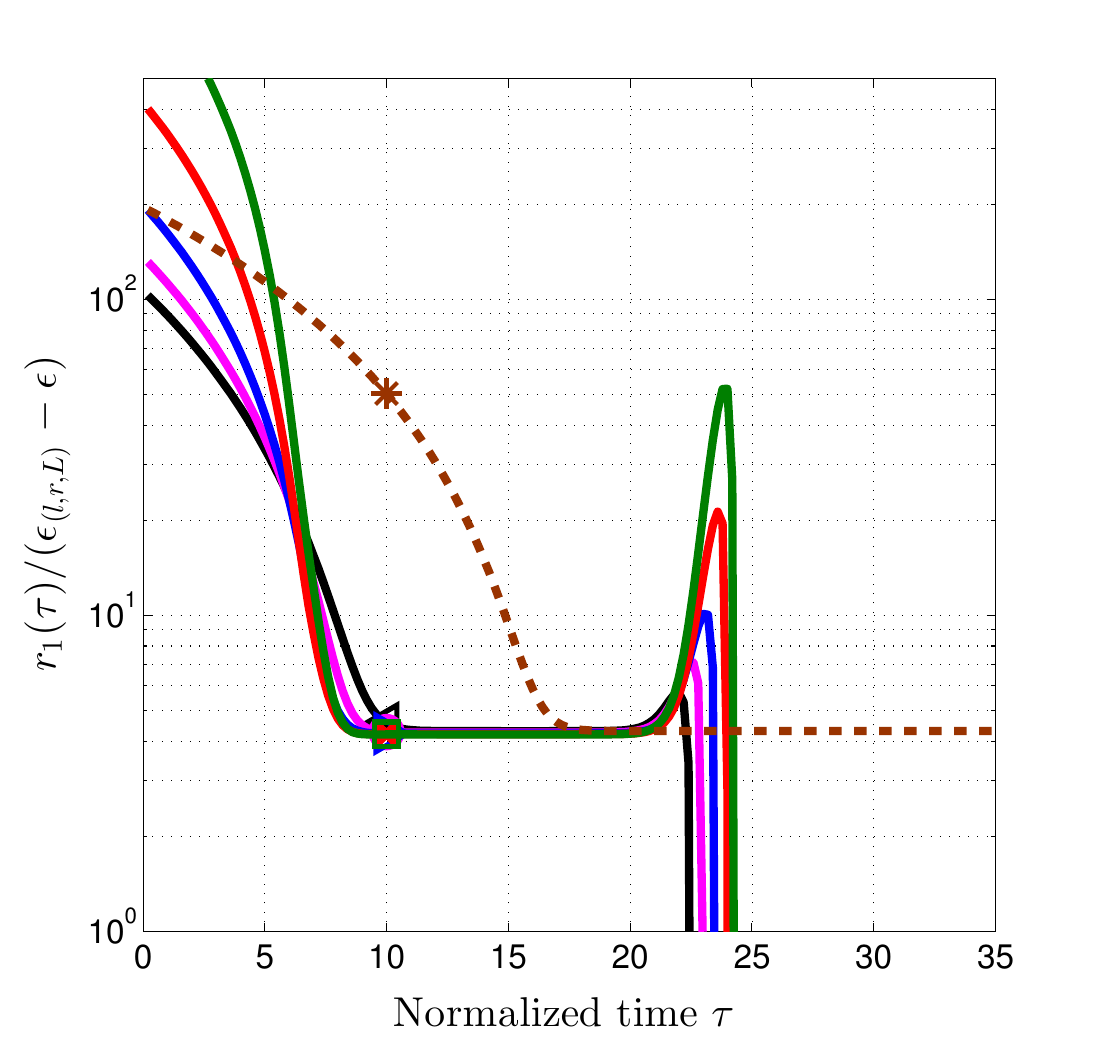}  & \includegraphics[scale=0.45]{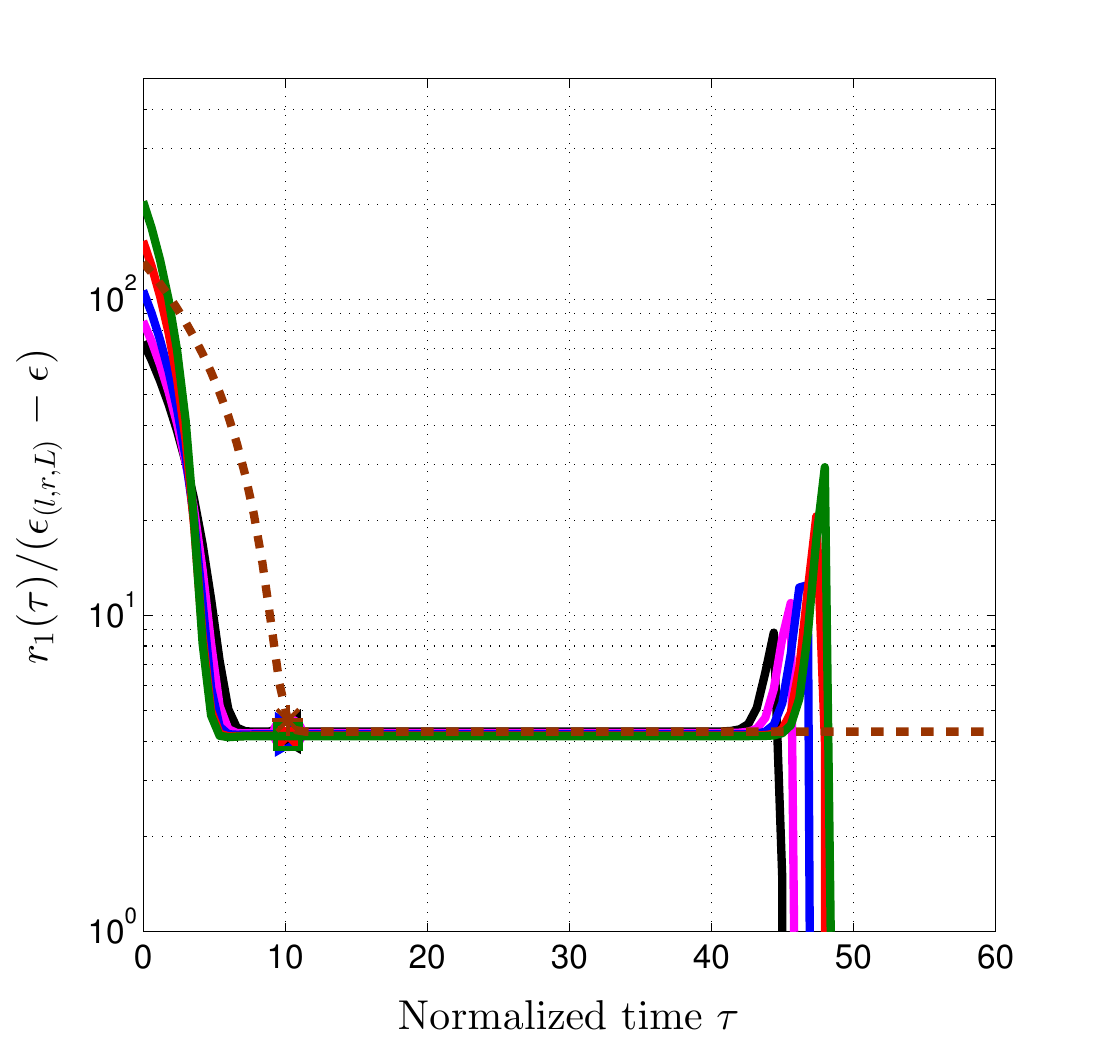}  \\
(a) & (b)
\end{tabular}
\caption{We plot $\hat{r}_{1}(\tau)/(\pe_{(l,r,L)}-\pe)$ for the  $(3,6,50)$ ensemble (a) and for the $(4,8,100)$ ensemble (b). In both figures, the $\pe$ considered are:  $0.45$ $(\lhd)$, $0.46$ $(\circ)$, $0.47$ $(\rhd)$, $0.48$ $(\times)$ and $0.485$ $(\square)$. The thresholds are respectively $\pe_{(3,6,50)}=0.48815$ and $\pe_{(4,8,100)}=0.49774$.  For the same ensembles and $\epsilon=0.45$, we also include $\hat{r}_{1}(\tau)/(\pe_{(l,r,L)}-\pe)$ for the double chain length, namely $L=100$ and $L=200$ respectively, dashed  lines with $(\ast)$ marker.
 } \LABFIG{FigDE2} \end{figure*}

The second phases starts at  $\tau^*$ and visually it corresponds
to two ``decoding waves'' that travel at constant constant speed
from the boundaries towards the center of the graph
\cite{KudekarWhy,Olmos11-3}.  In this phase the degree-one check
nodes that are being removed occur mostly around the position where
the decoding wave has its rapid rise. This can be seen in Fig.
\FIG{ProfilePU} by observing the  $p_u(\tau)$ and $\hat{v}_u(\tau)$
profiles for the $(3,6,50)$ ensemble at $\tau=15$.  Note that in
this second phase we do not have one critical time point at which
the decoder is most likely to stop, but the expected number of
degree-one check nodes is essentially a constant throughout this
critical phase.   Therefore, we call this the ``steady state'' phase.\footnote{As
pointed out in \cite{KudekarWhy}, the evolution is not completely
flat but exhibits small wiggles.  But these wiggles are extremely
small and, further, their amplitude vanishes as $l$ tends to infinity.
For instance, in Fig. \FIG{FigDE}(a), the amplitude of the oscillation
is $10^{-7}$.} Note that the higher we pick $\pe$ the closer
$\hat{r}_{1}(\tau)$ gets to the zero value. Actually, the $(l,r,L)$
threshold is given by the maximum $\pe$ value for which
$\hat{r}_{1}(\tau)>0$ during the steady state phase.  For the
$(3,6,50)$ ensemble the BP threshold is $\pe_{(3,6,50)}=0.48815$
and for the $(4,8,100)$ ensemble we get $\pe_{(4,8,100)}=0.4977$
\cite{Lentmaier10,KudekarWhy}.

It is important at this point to emphasize that we are looking at
a spatially-coupled LDPC code ensemble that is terminated at both
ends. During the decoding process there is both a decoding wave
that moves from the left end towards the middle as well as a decoding
wave that moves from the right end towards the middle. Due to these
{\em two} decoding waves, the expected number of degree-one check
nodes is {\em twice} what we would get if we considered an ensemble
that is terminated only at a single side. The same observation
applies to other quantities as well, e.g., the variance of the
number of degree-one check nodes  that we will compute soon. Rather
than aggregating the quantities corresponding to the two waves, we
could alternatively think of the decoding process as two processes
(the ``left'' process and the ``right'' process) and compute the
quantities for each of the individual processes separately.

From now on we will stick with the chosen model and leave it to the
reader to note the slight modifications that would be necessary if
we were to consider ensembles with a one-sided termination only.

\vspace{0.1cm}
\subsubsection{Third phase}
Finally, when the two decoding waves starting at the boundaries 
meet in the middle of the chain, a third phase
takes over. Since the expected fraction of degree-one check nodes
in the residual graph at both the first and third phase is significantly
larger than in the steady state phase, for most codes in the $(l,r,L)$
ensemble it is very unlikely for the decoder to declare a failure in either
the first or the last phase. Getting stuck in the first phase 
corresponds to an atypical erasure pattern, in which the  fraction
of erased bits at the boundary is significantly larger than $\pe$. This is an unlikely event however,
since we know that likely deviations are of the order of the standard deviation and it is proportional to 
$M^{1/2}$. In the last phase, most variable nodes have
already been decoded and still the graph contains a large number
of degree-one check nodes. For instance, for the $(3,6,50)$ ensemble
at $\pe=0.45$ and $\tau=22$, we can compute that the expected number
of variable nodes in the residual graph is approximately $0.31M$
while the number of degree-one check nodes is $0.202M$. Hence, a
large fraction of variable nodes can still be decoded. Errors in
this regime are typically caused by small cycles in the graph, or
stopping sets \cite{Urbanke08-2}. As shown in
\cite{Costello14,Mitchell11-2,Sridharan07}, SC-LDPC code ensembles have
linear growth of minimum distance with the block length $\n=ML$ and
thus codes with no low-weight stopping sets can be easily found for
sufficiently large $M$.

In the light of the above, we concentrate on the intermediate steady
state phase and inquire how we can express the error probability
as a function of the properties of the $r_1(\tau)$ random process
during this phase.

\subsection{A closer look at the steady state phase}\LABSEC{gamma}
Denote by $\hat{r}_{1}(*)$ the expected fraction of
degree-one check nodes during the steady state phase.  As expected,
$\hat{r}_{1}(*)\rightarrow 0$ as the channel parameter approaches
the BP threshold. In order to relate the $(l,r,L)$ average block
error probability to $\pe$, it will be convenient to find the correct 
scaling of the mean and the variance of the process
$r_1(\tau)$ in \EQ{r1} as a function of $\pe$. Following \cite{Urbanke09},
we consider a first-order Taylor expansion of both quantities
around the ensemble threshold $\pe_{(l,r,L)}$. For the case of
$\hat{r}_{1}(*)$, we get
\begin{align}\LABEQ{meanexpansion}
\hat{r}_{1}(*)|_{\pe}\approx\hat{r}_{1}(*)|_{\pe_{(l,r,L)}}+\gamma~ (\pe_{(l,r,L)}-\pe)+\mathcal{O}((\pe_{(l,r,L)}-\pe)^2).
\end{align}
Since $\hat{r}_{1}(*)|_{\pe_{(l,r,L)}}=0$ by definition, the $\gamma$ constant can be estimated from the numerical solution to $\hat{r}_{1}(*)$ for a given $\pe<\pe_{(l,r,L)}$ by
\begin{align}\LABEQ{gamma}
\gamma\approx \frac{\hat{r}_{1}(*)|_{\pe}-\hat{r}_{1}(*)|_{\pe_{(l,r,L)}}}{\pe_{(l,r,L)}-\pe}=\frac{\hat{r}_{1}(*)|_{\pe}}{\pe_{(l,r,L)}-\pe}.
\end{align}

In Fig.~\FIG{FigDE2}, we plot $\hat{r}_{1}(\tau)/(\pe_{(l,r,L)}-\pe)$
for the two ensembles considered in Fig.~\FIG{FigDE}. Note that
this quantity is essentially identical for all $\pe$ values in the
steady state phase. This indicates that indeed we can ignore the
quadratic and higher-order terms in \EQ{meanexpansion} close to the
threshold.  Hence, we assume the following scaling for the expected
fraction of degree-one check nodes during the critical phase
\begin{align}\LABEQ{mean}
\hat{r}_{1}(*)\approx \gamma (\pe_{(l,r,L)}-\pe).
\end{align}
Further, Fig.~\FIG{FigDE2} confirms, that, as expected, the constant
$\gamma$ is the same for the $(l,r,L)$ ensemble and the same ensemble
with twice the chain length, i.e., $(l,r,2L)$.  In Table\TAB{TABLAGAMMA},
we collect the values of $\gamma$ for various values of $l$ and $r$
computed for $L=100$. Also included are the MAP thresholds of the
underlying uncoupled  $(l,r)$-regular LDPC code ensemble, which are
up to numeric precision equal to the threshold $\pe_{(l,r,L)}$ for
the chosen value of $L$. In all cases, $\gamma$ is computed by
evaluating \EQ{gamma} at $\pe=\pe_{(l,r,L)}-0.04$.  \footnote{In
principle, we should compute these quantities as close as possible
to the threshold. But since we are using numerical integration
techniques, it is more stable to consider an $\pe$ value that is
further away from the threshold of the ensemble.  This is particularly
important for the integration of the covariance evolution equations
described in Section\SEC{COVPD}.  This motivated us to choose
$\pe=\pe_{(l,r,L)}-0.04$ as reference $\pe$ value to compute the
different scaling parameters.}

\begin{table}[h]
\begin{center}
\scalebox{0.9}{%
\begin{tabular}{|c|c|c|c|}\hline
l & r & MAP threshold & $\gamma$ \\\hline
$3$ & 6 & $0.4881$ & $4.31$\\
$4$ & $8$ & $0.4977$ & $4.24$\\
$5$ & $10$ &  $0.4994$ & $4.19$\\
$6$ & $12$ &  $0.4999$ & $4.15$\\
$4$ & $12$ &  $0.3302$ &  $4.28$ \\
$5$ & $15$ & $0.3325$ & $4.23$\\
$4$ & $6$ &  $0.6656$ &  $4.2$ \\
\hline
\end{tabular}}
\end{center}
\caption{$\gamma$ parameter for different $(l,r,L)$ ensembles, all with $L=100$.}\LABTAB{TABLAGAMMA}
\end{table}

So far we have seen how to determine $\gamma$ once we solved the
differential equation describing the peeling decoder. But $\gamma$
can also be determined via density evolution since it is closely
connected to the ``speed'' of the BP decoder \cite{KudekarWhy}, a further quantity
of significant practical importance. More precisely, consider the
following question.  Run a message-passing BP decoder with parallel updates and
consider again the ``steady phase'' of the decoder. How many
iterations\footnote{In general it will not take an integral
number of iterations to move by exactly one position but rather
this number will be fractional.} does it take until the decoding wave has ``moved'' by
one position?  It is natural to define the {\em
speed} of the decoder to be the inverse of this number.  Clearly,
the smaller the number of required iterations (the higher the speed)
the less complex the decoding is. The number of required iterations
is a function of the channel parameter and some thought shows that
it behaves like $c/(\pe_{(l,r,L)}-\pe)$ close to the threshold
\cite{Vahid2013}, where $c$ is a real positive constant.

\begin{figure}[h]
\centering \includegraphics[scale=0.5]{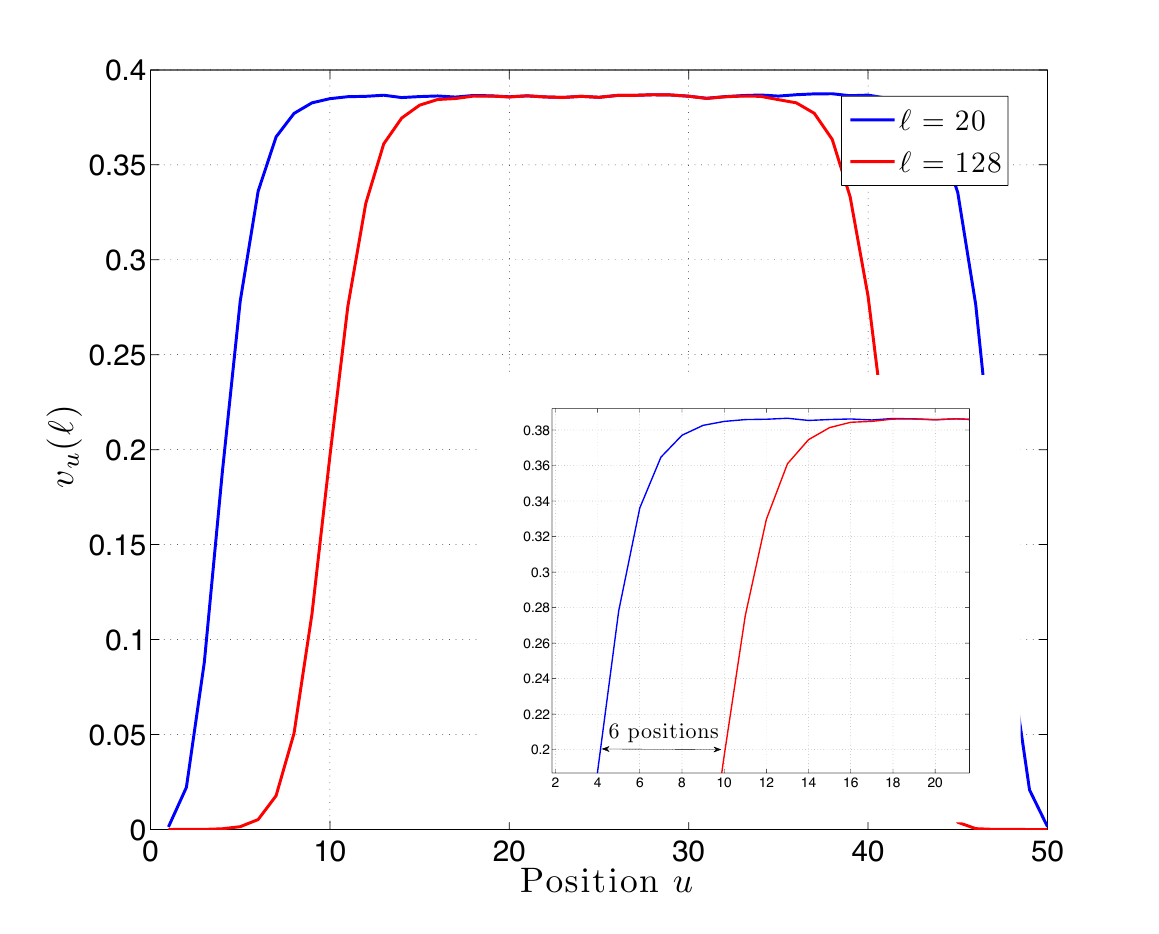}
\caption{For a $(3,6,50)$ code with $M=16000$ bits per position,
we plot the simulated profile of the fraction of variable nodes left in the
graph per position  $v_u(\ell)$, $u=1,\ldots, L$, after $\ell=20$ and
$\ell=128$ iterations of the parallel BP decoder. The channel parameter is 
$\pe=\pe_{(l,r,L)}-0.01$. Note that it takes about $128-20=108$ BP iterations
for the wave to move by $6$ positions.
}\LABFIG{wave36}
\end{figure}

\begin{figure*}[bh!]
\centering
\begin{tabular}{cc}
\includegraphics[scale=0.45]{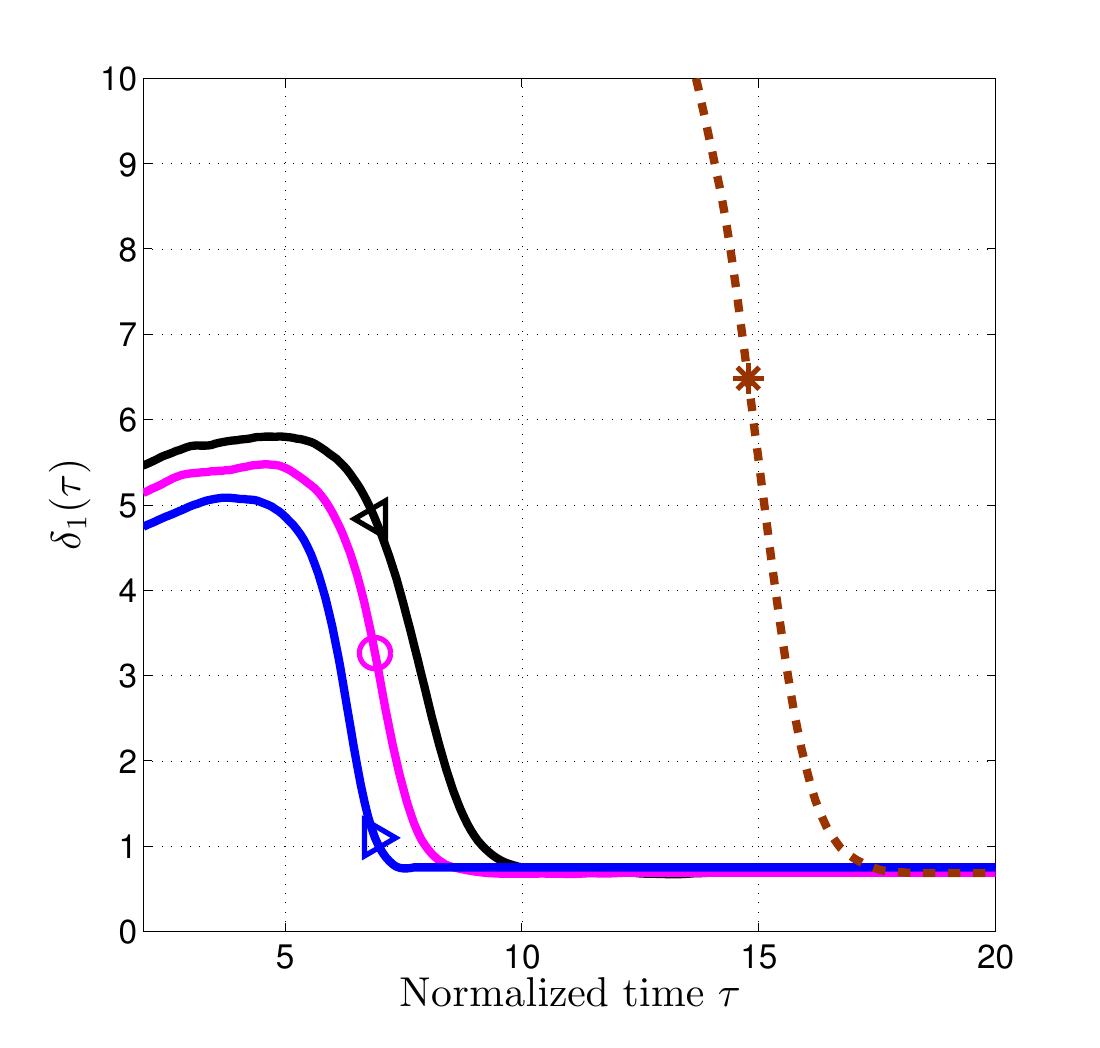}  & \includegraphics[scale=0.45]{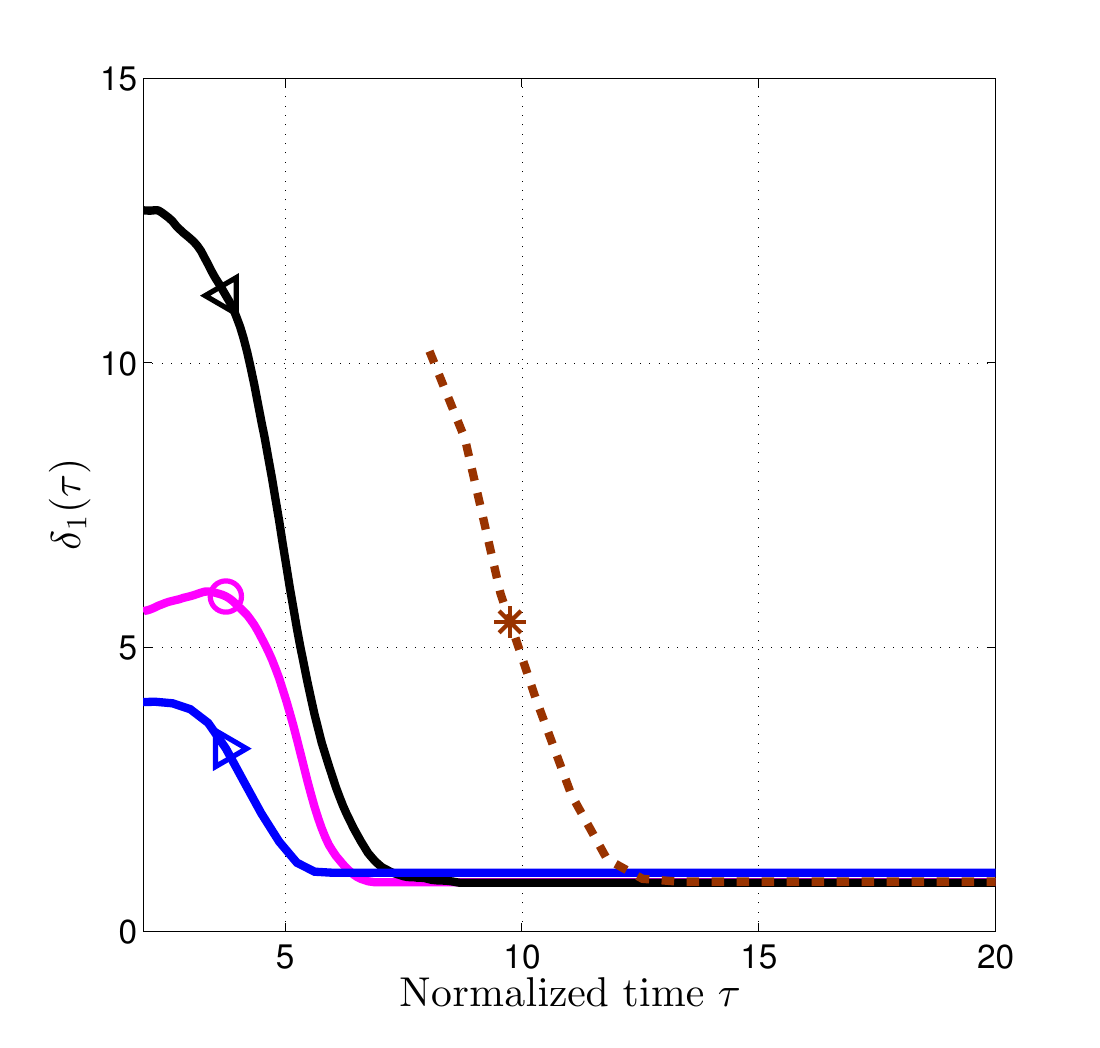}  \\
(a) & (b)
\end{tabular}
\caption{In (a), we plot $\delta_1(\tau)$
computed for the $(3,6,50)$ ensemble and  $\epsilon=0.45$ $(\lhd)$, $\epsilon=0.46$ $(\circ)$ and $\epsilon=0.47$ $(\rhd)$. In (b), we reproduce the same results for the $(4,8,100)$ ensemble. In dashed line with $*$ marker, we also plot $\delta_1(\tau)$   for  $L=100$ (a) and  for  $L=200$ (b) for $\pe=0.45$.}\LABFIG{FigVAR}
\end{figure*}

Assume that the decoding wave has moved by exactly one position
under the appropriate number of BP steps.  How many variable nodes
have we determined during these steps? The answer is $2 \beta M$,
where $\beta$ is the fraction of yet undetermined variables in the
uncoupled system when we transmit at parameter $\pe$ and
perform an infinite number of iterations (this is the fixed point
that we get stuck in when transmitting using an uncoupled code above
its BP threshold; the example below will hopefully clarify the exact
definition of $\beta$).  The factor $2$ stems from the fact that
we are looking at a spatially-coupled ensemble where both ends are
terminated and so there is both a ``left'' wave moving to the right
and a ``right'' wave moving to the left.

Next, note that if $\pe_{(l,r,L)}-\pe$ is very small then, with high probability, for each
variable that we determine there will be exactly one degree-one check node that is connected to it \cite{Vahid2013}.
%there will with  high probability correspond
%exactly one degree-one check node that is connected to it.
 Finally, by our definition, both the number of check nodes and the number
of variable nodes are normalized by $M$. It follows that we must have $\gamma=2\beta/c$. We note that
\cite{CFMZ13} shows how to compute the speed of the wave for a
spatially-coupled Curie-Weiss model and that \cite{Vahid2013} gives a
way to bound the speed of the wave for a particular spatially-coupled LDPC
code ensemble.  Both of these computations are in terms of quantities
that appear in density evolution.  It is therefore in principle
possible to compute the speed of the wave, and hence $\gamma$, for the $(l,r,L)$ ensemble based
on quantities that appear in density evolution.

\begin{example}\label{exa:gamma}
Let us look at one of our running examples, namely the $(3, 6, 50)$
ensemble. Rather than computing the decoding speed analytically,
let us determine the speed via simulations. From Fig.~\FIG{wave36}
we see that it takes about 108 iterations for the decoding wave to
move $6$ positions, where $\pe=\pe_{(l,r,L)}-0.01$. Therefore,
$c=\frac{108}{6} \cdot 0.01=0.18$.  To determine $\beta$, note that
for the $(3, 6)$ ensemble the BP message-passing decoder gets stuck in the point $x$
when transmitting at parameter $\pe$, where
$\pe(x)=\frac{x}{(1-(1-x)^5)^2}$ \cite{Urbanke08-2}.  With $\pe=\pe_{(l,r,L)}-0.01
=0.47815$ we get $x=0.41475$. We then have $\beta=\pe
(1-(1-x)^5)^3=0.386273$. Therefore, $\gamma=2 \beta/c=4.29$, which is a very good match to the 
$\gamma$ value in Table\TAB{TABLAGAMMA}.
\end{example}

We will soon see how $\gamma$ enters in the formula for the error
probability but the basic idea is simple. The larger $\gamma$ the
more degree-one check nodes we have in expectation at a given value
of the channel parameter. Recall that an error occurs if at any
point before the whole graph has not been peeled off we run out of
degree-one check nodes. In other words, an error occurs if the
actual number of degree-one check nodes deviates from the mean and
takes on the value zero.  So all other parameters (in particular
the variance) being equal, the larger the mean, the less likely
this event will be.

\begin{figure*}[b]
\hrulefill
% ensure that we have normalsize text
\normalsize
\begin{align}
&\text{Covariance evolution differential equations:}\nonumber\\\nonumber\\\LABEQ{deltaderiv} \tag{19.B}
&\frac{\partial \delta^{j,u}_{z,x}(\tau)}{\partial \tau}=f^{j,u}_{z,x}\left(\hat{\G}(\tau)\right)-f_{j,u}\left(\hat{\G}(\tau)\right)f_{z,x}\left(\hat{\G}(\tau)\right)+\sum_{q=1}^{r}\sum_{m=1}^{D}\delta^{j,u}_{q,m}(\tau)\frac{\partial f_{z,x}\left(\mathcal{G}(\tau)\right)}{\partial \G_{q,m}(\tau)}\Big|_{\hat{\G}(\tau)}+\delta^{q,m}_{z,x}(\tau)\frac{\partial f_{j,u}\left(\mathcal{G}(\tau)\right)}{\partial \G_{q,m}(\tau)}\Big|_{\hat{\G}(\tau)}\\\nonumber\\\nonumber
& \text{ for }(u,x)\in[D]^2, ~(j,z)\in[r+1]^2. ~ \hat{\G}(\tau) \text{ is the expected graph evolution computed by solving \EQ{system1}}.
\end{align}
% Restore the current equation number.
%\setcounter{equation}{\value{mytempeqncnt}}
% IEEE uses as a separator
\hrulefill
% The spacer can be tweaked to stop underfull vboxes.
\vspace*{4pt}
\end{figure*}

\subsection{A lower bound on $\tau^*$}\LABSEC{lower}

The time $\tau^*$ at which the steady state phase takes over is a
function of the uncoupled $(l,r)$-regular LDPC code ensemble,
the coupling pattern, the chain length $L$ and the channel parameter
$\epsilon$. A lower bound on $\tau^*$,  that we denote by
$\underline{\tau}$, can be obtained by ignoring the low-rate
terminations at both sides of the $(l,r,L)$ code graph and by
assuming a  $(l,r)$-regular LDPC code ensemble of length $\n=ML$
operating above its BP threshold. The bound $\underline{\tau}$ is then computed as the
expected time at which the graph runs out of degree-one
check nodes. This is essentially the same computation that we have performed
in Example~\ref{exa:gamma}.
\begin{example}
Let us look again at one of our running examples, namely the $(3, 6, 50)$
ensemble. As we have seen, when the uncoupled decoder transmits above the BP threshold
then the decoder gets stuck before all bits have been decoded and we have denoted
the fraction of undecoded bits by $\beta$.

Let us compute $\beta$ when $\pe=0.48815$. In this case the fixed point $x$ is equal to
$x=0.432261$ and $\beta=\pe (1-(1-x)^5)^3=0.406764$ \cite{Urbanke08-2}. Right after the initialization the
fraction of uncoded bits is $\epsilon$. Therefore the total number of bits that will
have been decoded in a chain of length $L$ (after division by $M$) is equal to $\underline{\tau}\doteq L(\pe-\beta)$.
For the particular case this gives us $L(\pe-\beta)=50(0.48815-0.406764)=4.0693$.
\end{example}

In Table~\TAB{TABLAtau} we state these bounds for a few examples in the form\footnote{As we mentioned, at the end of the decoding process, when the two decoding waves ``meet'', we enter
the third phase of the decoding process and during these phase decoding errors are again very unlikely. To get a better approximation to the error probability, one should also try to account for the length of this phase. } $\underline{\tau}/L$. For each $(l,r,L)$ configuration, $\beta$ is computed at $\epsilon_{(l,r,L)}$.
\begin{table}[t]
\begin{center}
\scalebox{0.9}{%
\begin{tabular}{|c|c|c|}\hline
l & r &  $\underline{\tau}/L$ \\\hline
$3$ & $6$ & $0.0814$ \\
$4$ & $8$ & $0.0193$  \\
$5$ & $10$ & $0.0053$  \\
$6$ & $12$ &  $0.0015$ \\
$4$ & $12$  &  $0.020$ \\
$5$ & $15$ & $0.0067$ \\
$4$ & $6$  &  $0.01272$  \\ 
\hline
\end{tabular}}
\end{center}
\caption{$\underline{\tau}/L$  for some regular ensembles.}\LABTAB{TABLAtau}\vspace{-4mm}
\end{table}

\section{Covariance Evolution}\LABSEC{COVPD}
Our next goal is to study the second-order statistics of the
$r_1(\tau)$ process during the steady state phase, which is the main
purpose of this section. As shown in \cite{Urbanke09}, the covariance evolution for the $(l,r,L)$ ensemble, i.e., the 
evolution along peeling decoding of moments of the form
$\text{CoVar}[\G_{j,u}(\tau),\G_{z,x}(\tau)]$ for any pair  of positions $(u,x)\in[D]^2$ and any pair of degrees $(j,z)\in[r+1]^2$, can
be estimated by solving an augmented system of differential equations, referred to as covariance evolution. Define
\begin{align}\LABEQ{Corr}
&f^{j,u}_{z,x}\left(\mathcal{G}(\tau)\right)=\\
&\frac{\E\left[\Big(\G_{j,u}(\tau+\frac{1}{M})-\G_{j,u}(\tau)\Big)\Big(\G_{z,x}(\tau+\frac{1}{M})-\G_{z,x}(\tau)\Big)\Big| \mathcal{G}(\tau) \right]}{1/M^2}.\nonumber
\end{align}

If the system of differential equations in \EQ{system1} is augmented with the  set of equations in \EQ{deltaderiv}, then the solution for $\delta^{j,u}_{z,x}(\tau)$ is also unique and   given the initial conditions
\begin{align}\LABEQ{COV1}
\hat{\G}_{j,u}(0)&=\E[\G_{j,u}(0)],\\\LABEQ{COV2}
\delta^{j,u}_{z,x}(0)&=\E[\G_{j,u}(0)\G_{z,x}(0)]-\E[\G_{j,u}(0)]\E[\G_{z,x}(0)],
\end{align}
the difference with respect the true covariance  $\text{CoVar}[\G_{j,u}(\tau),\G_{z,x}(\tau)]$ at any time $\tau$ is less than $M^{-1/2}$ \cite{Urbanke09}. Further, in the limit $M\rightarrow\infty$ the following holds
\begin{itemize}
\item $\G_{j,u}(\tau)$ is Gaussian distributed with mean $\hat{\G}_{j,u}(\tau)$ and variance $\delta^{j,u}_{j,u}(\tau)/M$. 
\end{itemize}
\begin{itemize}
\item For any pair  of positions $(u,x)\in[D]^2$ and any pair of degrees $(j,z)\in[r+1]^2$, $\G_{j,u}(\tau)$ and $\G_{z,x}(\tau)$ are jointly Gaussian distributed with cross covariance $\delta^{j,u}_{z,x}(\tau)/M$.
\end{itemize}

%
%Further,  $\G(\tau)$ converges (in $M$) to a 
%multivariate Gaussian distribution of dimension $D\times(r+1)$ with mean $\hat{\G}(\tau)$
%and  covariance matrix defined by   $\delta_{ju,zx}(\tau)/M$ for $j,z\in\{1,\ldots,r+1\}$ and $u,x\in\{1,\ldots,D\}$. 
Note that any covariance moment vanishes in the limit $M\rightarrow\infty$ and hence $\G(\tau)$ concentrates around the mean $\hat{\G}(\tau)$ predicted by \EQ{system1}. The details of the computation of the correlations in \EQ{Corr} can be found in  Appendix\SEC{A2}. In Appendix\SEC{A3}, we derive the initial conditions in
 \EQ{COV2} for the $(l,r,L)$ ensemble.  The derivatives of the function $f_{j,u}\left(\mathcal{G}(\tau)\right)$  with respect to any component $\G_{q,m}(\tau)$, used  in \EQ{deltaderiv}, can be computed given the closed-formed expression to $f_{j,u}\left(\mathcal{G}(\tau)\right)$ derived Appendix\SEC{A1}. 

To estimate the  error probability of the $(l,r,L)$ ensemble, we need to evaluate the variance of the process $r_1(\tau)$ in \EQ{r1}, whose expected evolution has been described in Section \ref{S1}. In the light of the above results, for sufficiently large $M$  the distribution of $r_1(\tau)$ is well approximated by a Gaussian process whose variance can be obtained given the solution to the covariance evolution equations in \EQ{deltaderiv}:
\begin{align}\nonumber
\text{Var}[r_1(\tau)]&=\mathbb{E}[\big(r_1(\tau)-\hat{r}_1(\tau)\big)^2]
\\\LABEQ{var1}
&=\frac{1}{M}\displaystyle \sum_{u=1}^{D}\sum_{x=1}^{D}\delta_{1u,1x}(\tau)\doteq \frac{\delta_1(\tau)}{M}.
\end{align}

\begin{table}[b]
\begin{center}
\scalebox{0.9}{%
\begin{tabular}{|c|c|c|c|}\hline
l & r &  $\delta_1^*$ & $\gamma/\sqrt{\delta^*_{1}}$ \\\hline
$3$ & 6 & $0.67$ & $5.12$ \\
$4$ & $8$ & $0.85$ & $4.44$ \\
$5$ & $10$ & $0.91$ & $4.23$ \\
$6$ & $12$ &  $1.05$ & $4.04$\\
$4$ & $12$  &  $0.64$ & $5.1$ \\
$5$ & $15$ & $0.72$& $4.72$ \\
$4$ & $6$  &  $0.91$ & $4.28$ \\ 
\hline
\end{tabular}}
\end{center}
\caption{$\delta^*_{1}$  parameter computed for different $(l,r,L)$ ensembles.}\LABTAB{TABLANU}\vspace{-4mm}
\end{table}

In Fig. \FIG{FigVAR}, we plot the solution for $\delta_{1}(\tau)$
computed numerically for the $(3,6,50)$ ensemble (a) and for the
$(4,8,100)$ ensemble (b) for the following $\pe$ values: $0.45$
$(\lhd)$, $0.46$ $(\circ)$, and $0.47$ $(\rhd)$. The dashed lines
with the $(*)$ marker  correspond  to $\delta_{1}(\tau)$ for
$\pe=0.45$ and the same ensembles with twice the chain lengths,
i.e., $L=100$ in (a) and $L=200$ in (b). As $\hat{r}_{1}(\tau)$ in
Fig. \FIG{FigDE}, $\delta_1(\tau)$ presents an approximate
a flat evolution during the steady state phase (as in the mean
evolution case, small wiggles can be observed by magnification).
Denote by $\delta^*_{1}$ the  value of the variance during the
steady state phase. Note also that $\delta^*_{1}$  is roughly
constant with $\pe$ for the set of values considered. Indeed, this
same effect has been noticed for all tested   configurations of the
$(l,r,L)$ ensemble and $\pe$ values sufficiently distant  from the
ensemble threshold $\pe_{(l,r,L)}$. In the following, we will use
as representative value to  $\delta^*_{1}$ the one computed at
$\pe=\pe_{(l,r,L)}-0.04$. Obviously, further corrections on the
scaling law that is proposed in this paper to estimate the $(l,r,L)$
performance can consider dropping this assumption and modeling
$\delta^*_{1}$ as a function of $\pe$, or integrating the covariance
evolution equations for each $\pe$ value. In Table\TAB{TABLANU},
we summarize the  $\delta^*_{1}$ values computed for different
configurations of the $(l,r,L)$ ensemble with $L=100$.

An important figure of merit to evaluate the finite-length scaling
properties of  $(l,r)$-regular LDPC code ensembles is the ratio of the
expected number of degree-one check nodes at the critical point,
$\hat{r}_1(*)$, to its standard deviation \cite{Urbanke09}. As we show in Section\SEC{SL},
the same figure of merit, associated to the mean and variance of
the $r_1(\tau)$ process during the critical phase, appears in the
scaling law proposed for the $(l,r,L)$ ensemble:
\begin{align}\LABEQ{ratio}
\frac{\hat{r}_1(*)}{\sqrt{\delta^*_{1}/M}}\approx \frac{\gamma \sqrt{M}(\pe_{(l,r,L)}-\pe)}{\sqrt{\delta^*_{1}}}.
%=\alpha \sqrt{M} (\pe_{(l,r,L)}-\pe)
\end{align}
One of the main
conclusions of this work is that the error probability of the
$(l,r,L)$ decreases exponentially fast with \EQ{ratio}. Thus, small differences in $\gamma/\sqrt{\delta^*_{1}}$ for
different configurations might have a noticeable impact in the
finite-length performance of the code. In Table\TAB{TABLANU}, we
also include the  ratio $\gamma/\sqrt{\delta^*_{1}}$ computed for the different $(l,r,L)$ ensembles (
$\delta^*_{1}$ is  in the same table and $\gamma$ can be found in
Table\TAB{TABLAGAMMA}). Note that, for fixed $l/r$ ratio, $\gamma/\sqrt{\delta^*_{1}}$
tends to decrease as we increase $l$, suggesting that once the
threshold is sufficiently close to capacity, increasing the code
density does not improve the finite-length performance of the code.
For instance, this will be the case between the $(5,10,100)$ and
$(6,12,100)$ ensembles.

\subsection{Process covariance at two time instants}\LABSEC{CovTime}

%We are interested in estimating the zero-crossing probability of the process $r_1(\tau)$, which corresponds to a decoding failure.

As we mentioned before, unlike for the uncoupled $(l,r)$-regular LDPC code ensemble, the error
probability of the coupled ensemble is not determined by the behavior
of the decoder at a particular ``critical'' point in time.  As
illustrated in Fig. \FIG{FigDE}, the decoder remains in a critical
state during a period of time of  duration $\Theta(L)$.  Therefore,
the error probability in the coupled case is given by  the cumulative
probability that the $r_1(\tau)$ process hits zero at some point
during the critical phase. Recall that for the purpose of this
paper we ignore errors which may happen in either the initial or
the final phase, since they are very rare.

% the decoding process. Recall that for the purpose of this
%paper we ignore errors which may happen in either the initial or
%the final phase, since they are very rare and for our purpose the
%steady state phase is essentially the whole time except for some
%short period at the beginning and at the end.

To compute the error probability during the steady state phase, we have
to take into account the covariance of the $r_1(\tau)$ process over
time:
%
%
%%this error
%%probability is not determined by the behavior of the decoder at a
%%particular critical point in time, but errors happen more or less
%%uniformly throughout the whole steady state phase.  
%
%
%A first naive estimate is to assume that, provided the Gaussian distribution of $r_1(\tau)$ at the steady state phase $\tau$ with  known mean an variance, errors occur with  the same probability ant any time $\tau$ in this phase and that such probability is given as the probability of an univariate Gaussian with mean $\hat{r}_1(\tau^*)$ and variance $\delta_1(\tau^*)$ to take
%a negative value.
%%
%%given mean and variance
%%of the degree-one check nodes, and given that their distribution
%%is Gaussian, an error occurs with the same probability, as the
%%probability that a Gaussian with given mean and variance takes on
%%a negative value. 
%Such an estimate ignores the correlations along
%time and would therefore be considerably off. Let us therefore
%discuss how we can incorporate these correlations. More precisely,
%in addition to the the mean and variance of $r_1(\tau)$ at a particular time $\tau$, we need
%to estimate the process covariance along time, namely
\begin{align}\LABEQ{COV_TIME}
\phi_{1}(\zeta,\tau)\doteq\E[r_1(\zeta)r_1(\tau)]-\hat{r}_1(\zeta)\hat{r}_{1}(\tau),
\end{align}
where $\tau$ and $\zeta$ are two distinct time instances.  The
expectation is defined over the joint probability distribution of
the DD at times $\tau$ and $\zeta$, and we denote the corresponding
p.d.f.  by $p_{\G(\zeta),\G(\tau)}(g(\zeta),g(\tau))$. The quantity
$\phi_{1}(\zeta,\tau)$ can in principle be computed analytically
by a procedure similar to covariance evolution. However, while the
number of coupled equations in the covariance evolution system in
\EQ{deltaderiv} is $(D\times (r+1))^2$, the number of coupled
equations in the system that we would need to solve to be able to
compute $\phi_{1}(\zeta,\tau)$ analytically is at least $(D\times
(r+1))^4$ since we have to consider a minimum of two consecutive
PD iterations. This approach is complex and computationally
challenging.

Some thought shows that $\phi_{1}(\zeta,\tau)$ in \EQ{COV_TIME}
should be a function of $|\tau-\zeta|$ and that the decay of the
correlation should be exponential in this time difference, i.e.,
\begin{align}\LABEQ{corr} \phi_{1}(\zeta,\tau)\approx
\frac{\delta^*_{1}}{M} \text{e}^{-\theta|\tau-\zeta|}, 
\end{align}
where $\theta$  is a parameter that depends on the $(l,r)$-regular LDPC ensemble
and the coupling pattern of the SC-LDPC code.  Simulations support
the ansatz \EQ{corr}.  For instance, in Fig.~\FIG{COV_TIME}, we
plot (thin solid line) an estimation of $M\phi_{1}(\zeta,\tau)$ for
the $(3,6,50)$ ensemble, $\epsilon=0.45$ and $\zeta=13$ computed
by simulating $500$ transmitted codewords with a code generated
with $M=1000$ bits per position. Empirically we
have observed that in order to accurately estimate the value of
$\theta$ for a given $(l,r,L)$ ensemble, we only require
a few hundred transmitted codewords. Why is it natural to consider the ansatz in \EQ{corr}?  In \cite{KudekarMacris},
in the context of the finite-length analysis of uncoupled ensembles,
it was shown that in this case the correlation of BP messages decays exponentially
in the graph distance. For us the natural equivalent of graph distance
is the difference in decoding time.

\begin{figure}[h]
\centering \includegraphics[scale=0.45]{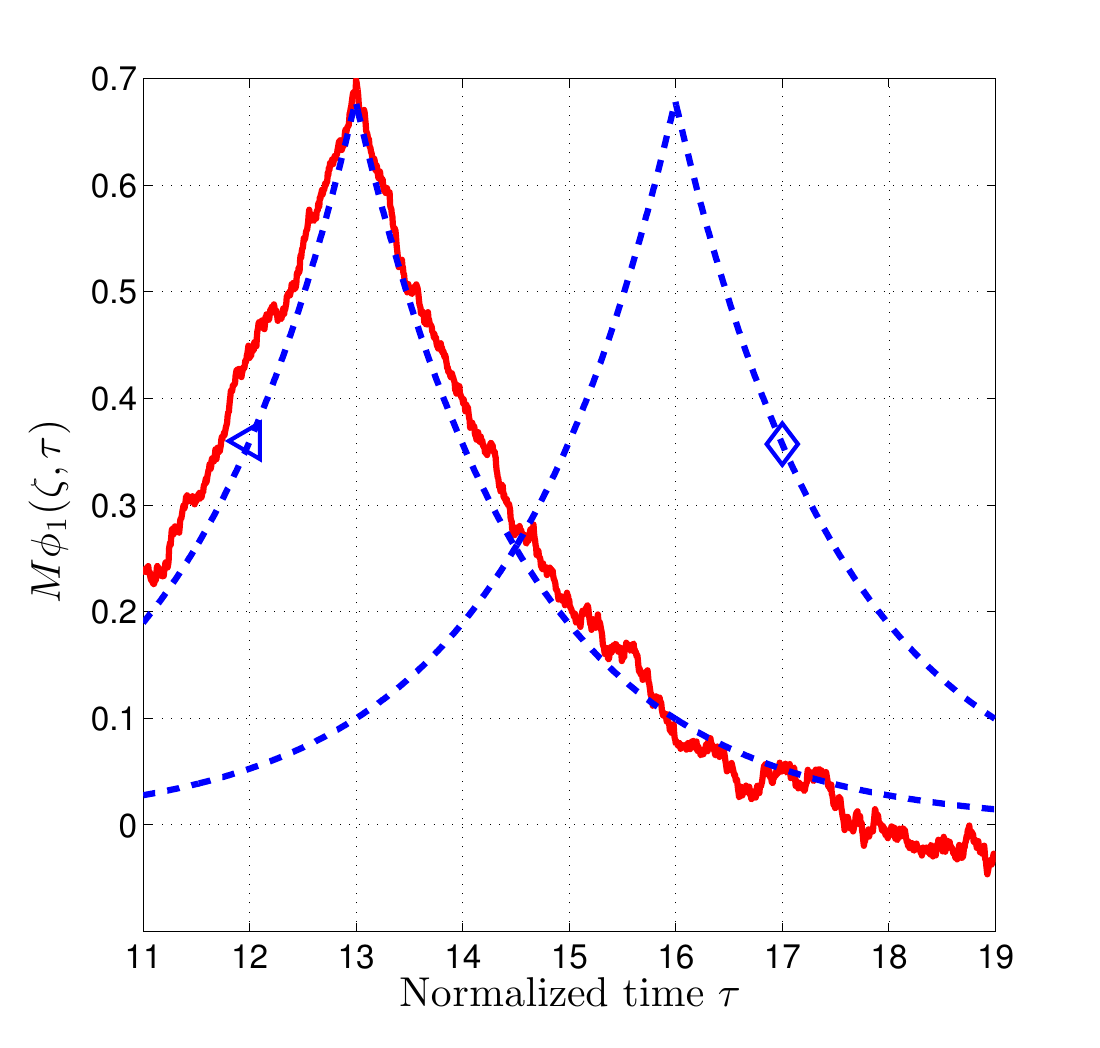}
\caption{$M \phi^{N}_{1}(\zeta,\tau)$ (dashed lines) computed for the $(3,6,50)$ ensemble and $N=200$ samples. We have fixed $\zeta$ to 13 $(\lhd)$ and 16  $(\diamondsuit)$. For $\zeta=13$, we have also included an estimation of  $M\phi_{1}(\zeta,\tau)$ computed using $500$ transmitted codewords for a code generated with $M=1000$ bits per position.}\LABFIG{COV_TIME}
\end{figure}

An alternative method to estimate $\theta$, that is based solely
on the analytical expressions derived before to compute the mean
and covariance of the residual graph along peeling decoding, is as follows. With
no loss of generality, we assume $\tau>\zeta$.  First, we express
the correlation term in \EQ{COV_TIME} in a more convenient way:
\begin{align}
&\E[r_1(\zeta)r_1(\tau)]\nonumber\\\nonumber
&=\int \int ~  r_1(\zeta) r_1(\tau)~ p_{\G(\zeta),\G(\tau)}(g(\zeta),g(\tau)) ~\text{d}g(\zeta)\text{d}g(\tau)\\
&=\int r_1(\zeta)~ \E[r_1(\tau)|g(\zeta)] ~ p_{\G(\zeta)}(g(\zeta)) ~\text{d}g(\zeta), \LABEQ{CORR_TIME}
\end{align}
where the expectation inside the integral in \EQ{CORR_TIME} is taken with respect the conditional probability distribution 
\begin{align}
p_{\G(\tau)| g(\zeta)}(g(\tau)),
\end{align}
i.e. the graph DD probability distribution at $\tau$ if the graph
DD at $\zeta$ is fixed to $g(\zeta)$. Note that, for any DD $g(\zeta)$, we
can easily  evaluate $\E[r_1(\tau)|g(\zeta)]$ using the same equations
derived to compute expected graph evolution in Section\SEC{PD} and
initial conditions given by $g(\zeta)$. Regarding the distribution
$p_{\G(\zeta)}(g(\zeta))$ of the DD at time $\zeta$, all the
$\G_{j,u}(\tau)$ terms in $\G(\tau)$, $u\in[D]$ and $j\in[r+1]$,
are jointly Gaussian distributed with  mean and covariance given
as the solution to the system of differential equations in \EQ{system12}
and \EQ{deltaderiv}.

Therefore, the solution to the integral in $\EQ{CORR_TIME}$ can be
estimated by taking samples\footnote{Sampling from $\G(\zeta)$ is
straightforward since it is Gaussian distributed.} from $\G(\zeta)$, where
each sample is a possible DD at $\zeta$. For each sample,
$\E[r_1(\tau)|g(\zeta)]$ can be obtained by numerical integration
of the expected graph evolution equations in \EQ{system12} using
the sampled DD as initial conditions.  Let $\mathcal{S}$ be the
collection for $N$ samples taken of the graph DD at time $\zeta$.
The integral in \EQ{CORR_TIME} is approximated by a sum of the form
\begin{align}
\frac{1}{N}\sum_{g(\zeta)\in \mathcal{S}} r_1(\zeta)\E[r_1(\tau)|g(\zeta)].
\end{align}
In the limit of $N\rightarrow\infty$ 
\begin{align}\LABEQ{COV_TIME2}
\phi^{N}_{1}(\zeta,\tau)\doteq\frac{1}{N}\sum_{g(\zeta)\in \mathcal{S}} r_1(\zeta)\E[r_1(\tau)|g(\zeta)]-\hat{r}_{1}(\zeta)\hat{r}_1(\tau)
\end{align}
converges to the true covariance $\phi_{1}(\zeta,\tau)$ \cite{Murphy}.
It is important to remark that $\phi^{N}_{1}(\zeta,\tau)$ is a
quantity obtained using exclusively the equations that  predict the
mean and variance graph evolution for the $(l,r,L)$ ensemble derived
in Sections\SEC{PD} and\SEC{COVPD} respectively.

In Fig. \FIG{COV_TIME}, we plot in dashed lines
$M\phi^{N}_{1}(\zeta,\tau)$ computed for the $(3,6,50)$ ensemble
at $\epsilon=0.45$ with $N=200$ samples. We have fixed $\zeta$ to
13 $(\lhd)$ and 16  $(\diamondsuit)$. As predicted,
the covariance decays exponentially with  $|\tau-\zeta|$.
%For this case, we obtain $\theta\approx 0.59$.
%\begin{align}\LABEQ{corr}
%\phi_{1}(\zeta,\tau)\approx \frac{\nu}{M} \text{e}^{-\theta|\tau-\zeta|},
%\end{align} 
%where $\theta\approx 0.59$ for this case. 
In Table\TAB{thetas}, we show the $\theta$ value computed for
different $(l,r,L)$ ensembles at $\pe=\pe_{(l,r,L)}-0.04$. As we
can observe, for fixed rate, the covariance decays slightly faster
as we increase  the check degree. A more detailed description of
the effect of $\theta$ (and the rest of scaling parameters) in the
finite-length performance is left to Section\SEC{SL}.
%An intuitive explanation is as follows. As the check node degree grows, 
%The higher the check node degree is,  the more weak the effect of the removal of a degree-one check node at a certain position. By weaker effect we mean that the probability of creating new degree-one  check nodes in nearby positions is smaller. 

\begin{table}[h]
\begin{center}
\scalebox{0.9}{%
\begin{tabular}{|c|c|c|}\hline
l & r &  $\theta$ \\\hline
$3$ & 6 & $0.59$\\
$4$ & $8$ & $0.61$\\
$5$ & $10$ & $0.63$\\
$4$ & $12$  &  $0.84$ \\
$5$ & $15$ & $0.88$\\
$4$ & $6$  &  $0.51$ \\ %0.98
\hline
\end{tabular}}
\end{center}
\caption{$\theta$ parameter for different $(l,r,L)$ ensembles.}\LABTAB{thetas}
\end{table}

Finally, it is important to note that the decay of correlation of the
$r_1(\tau)$ process, and hence the $\theta$ parameter in \EQ{corr},
does not only depend on the uncoupled $(l,r)$-regular LDPC code ensemble 
but also on the coupling pattern that we use to generate the coupled ensemble.
In order to illustrate this dependence, consider
a modification of the  $(l,r,L)$ ensemble, denoted by
$\mathcal{E}(l,r,L,w)$  where $w$ is a positive integer. For this
ensemble, the coupling pattern  as follows: for $u=1,\ldots,D$
\begin{itemize} \item If $u$ is odd, then each variable node at
position $u$ is connected to a check node at position $u,u+w,
u+2w,\ldots, u+w(l-1)$.  \item If $u$ is even, then each variable
node at position $u$ is connected to a check node at position $u,u+1,
u+2,\ldots, u+(l-1)$.  \end{itemize}
In words, we are ``stretching'' out the connections over a length
that is $w$ times larger. In analogy to convolutional codes, we
are increasing the ``constraint length'' of the code. Note that this ensemble has $D=L+w(l-1)$ positions and hence the
design rate is smaller than the design rate of the  $(l,r,L)$
ensemble.  A representation of the $\mathcal{E}(3,6,L,2)$ ensemble
can be found in Fig. \FIG{Expanded}.

What is the expected effect of this modification? As we increase $w$, we are further and further spreading out the
connections in the spatial dimension and hence we expect the slower
decay of the covariance of the process $r_1(\tau)$.
%as $w$ is increased,  the degree-one check nodes that
%are created after each PD iteration are spread in a larger region
%of the graph. Therefore, if $\tau_u$ and $\tau_x$ are the times at
%which the rapid rise of the decoding wave (see Fig. \FIG{ProfilePU})
%is approximately at position $u$ and position $x>u$ respectively,
%for larger $w$ it must be true that it is also higher the fraction
%of degree-one check nodes in the graph at $\tau_x$ that were created
%after removing those at $\tau_u$. Consequently, the covariance
%between $r_1(\tau_u)$ and $r_1(\tau_x)$ must grow with $w$.
Indeed, this effect can be observed in Fig.~\FIG{COV_TIMEW}, where
we show the covariance decay of the process $r_1(\tau)$ for the $(3,6,100)$ ensemble $(\lhd)$ and the
$\mathcal{E}(3,6,100,w)$ ensemble with $w=2$ $(\rhd)$ and $w=4$
$(\square)$. For each ensemble we represent
$M\phi_{1}(\zeta,\tau)/\delta^*_{1}$ so that all curves have a
maximum equal to $1$. The covariance has been estimated using $500$
transmitted codewords (thin dashed lines) and using the alternative
method summarized by equation \EQ{COV_TIME2} (solid lines), with
$N=200$  samples.  As expected, the covariance between $r_1(\tau)$
and $r_1(\zeta=29)$ is higher for larger $w$ values. The estimated
$\theta$ values are, respectively, $\theta=0.59$ , $\theta=0.28$
$(w=2)$ and $\theta=0.17$ $(w=4)$.

\begin{figure}[h]
\centering
\begin{tabular}{c}
\includegraphics[scale=0.65]{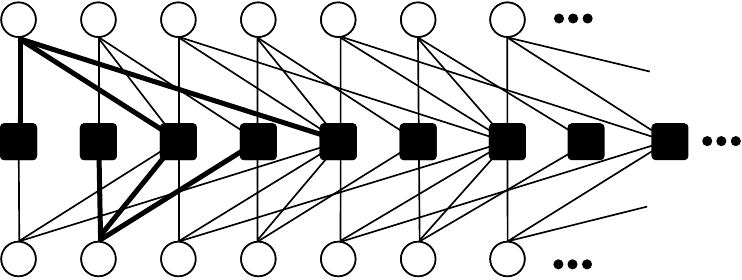}  
\end{tabular}
\caption{A graphical representation of the $(3,6,L,w)$ ensemble for $w=2$. In thick lines, we illustrate the connections of a variable node placed at an odd position and a variable node at an even position.}\LABFIG{Expanded}
\end{figure}

\begin{figure}[h]
\centering \includegraphics[scale=0.45]{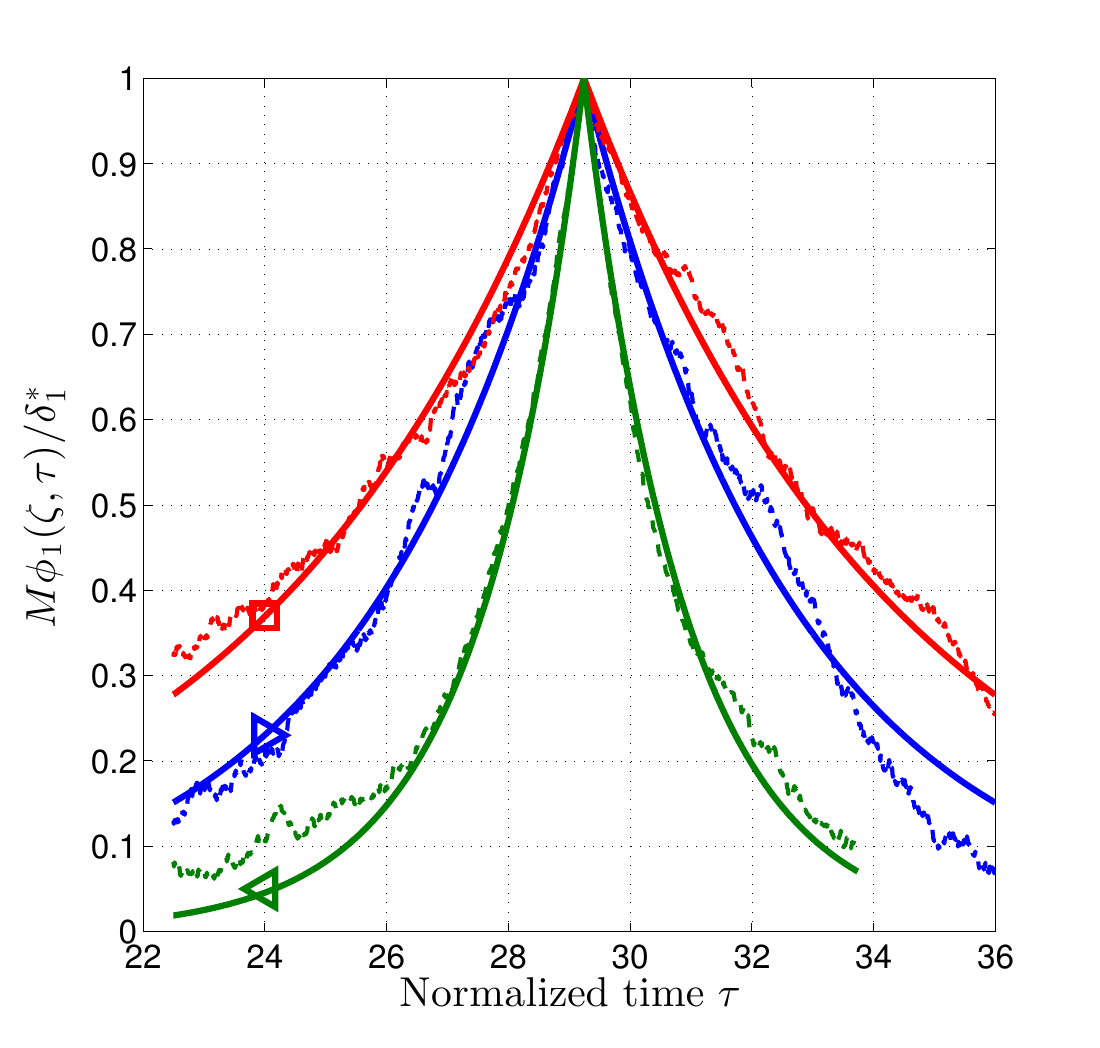}
\caption{$\phi^{N}_{1}(\zeta,\tau)$ with $N=200$ samples computed for the  $(3,6,100)$ ensemble $(\lhd)$ and the $\mathcal{E}(3,6,100,w)$ ensemble with $w=2$ $(\rhd)$ and $w=4$  $(\square)$. We have fixed $\zeta$ to 29. In thin dashed lines, we also include the corresponding estimate  computed using $500$ transmitted codewords. }\LABFIG{COV_TIMEW}
\end{figure}

At this point, we have all the ingredients we need  to predict the
survival probability of the process $r_1(\tau)$. Before concluding
the present section, we finally want to emphasize that the triple
$(\gamma, \delta_1^*, \theta)$ only depends on the uncoupled ensemble
 and the coupling pattern. The larger we choose $L$, the
longer the process $r_1(\tau)$ remains in the steady state phase
and thus the larger the error probability will be.  However, the
intrinsic statistical properties of the $r_1(\tau)$ during this phase process
do not vary if we increase the chain length $L$. To illustrate this
property, in Fig.~\FIG{FigDE} and Fig.~\FIG{FigVAR} we have  included
results  for both the  $(3,6,50)$ and $(4,8,100)$ ensembles and the
same ensembles with double chain lengths, $L=100$ and $L=200$
respectively.

\section{Prediction of the error probability using Ornstein-Uhlenbeck processes}\LABSEC{SL}

In  previous sections, we have provided  a statistical
characterization of the process $r_1(\tau)$ for the $(l,r,L)$
ensemble when used for transmission over the BEC and decoded via
PD. As proven in \cite{Urbanke09}, $r_{1}(\tau)$ is a Markov process
that converges (in the number $M$ of bits per position of the code)
to a Gaussian process. Further, covariance evolution for the $(l,r,L)$
ensemble shows that $r_1(\tau)$ in the  steady state phase is
essentially a  constant-mean and constant-variance process whose
temporal covariance only depends on the time difference between the
two time points considered.  In other words,  in the steady state
phase, $r_{1}(\tau)$ is well modeled by a stationary Gaussian Markov
process.
%Further, this forces the covariance to exponentially decay in the time difference \cite{OU-book}, which  is consistent with
%$\phi_{1}(\tau,\zeta)$ in \EQ{corr}. 
%In the light of these results, the evaluation of the $(l,r,L)$ decoding error
%probability is intimately linked with the first-passage time
%statistics of stationary Gaussian Markov processes.  
Indeed, a stationary Gaussian Markov process $X(t)$ can only be one
of the two following types \cite{OU-book, Doob42}:
\begin{itemize}
\item[a)] If $t_1<t_2<\ldots<t_n$, $X(t_1),\ldots,X(t_n)$ are mutually-independent Gaussian random variables.
\item[b)] There exists a constant $\alpha$ such that if
$t_1<t_2<\ldots<t_n$, then $X(t_1),\ldots,X(t_n)$ are jointly
distributed by a multivariate Gaussian with common mean and variance
and covariance function $\text{CoVar}[X(t+T),X(t)]\propto \exp^{-\alpha
T}$,
\end{itemize}
where the latter type of Gaussian Markov process is called an
Ornstein-Uhlenbeck (OU) process \cite{OU-book}. Therefore, under
the Gaussian assumption for the distribution of $r_1(\tau)$, an OU process is the only type of stationary
Gaussian Markov process that is compatible with the process
$r_1(\tau)$.
%
%
%Based on the first and second
%moments computed for $r_{1}(\tau)$ in former sections, an OU process emerges as a convenient model to predict the survival probability of the decoding
%process $r_1(\tau)$.  
In the following we describe the properties and parameters of OU
processes and link them with those already computed for $r_1(\tau)$
in previous sections.

\subsection{Ornstein-Uhlenbeck Processes}

An OU process is a stationary Gaussian Markov process evolving via
the following stochastic equation \cite{OU-book}:
\begin{align}
X(t)=X_0\text{e}^{-a t}+\sqrt{2b}\text{e}^{-a t}\int_{0}^ {t}\omega(u)\text{e}^{a u} \text{d}u, 
\end{align}
where $\omega(t)$ is a white noise with zero mean and unit variance,
$a$ and $b$ are real positive constants and $X_0$ is the initial condition.
Consider samples uniformly taken from $X(t)$ every $\Omega$ seconds,
where $X_i$ is the $i$-th sample.  The mean of $X_i$ is given by
\begin{align}
\E[X_i]=X_0\text{e}^{-a i\Omega}=X_0 g^{i}, \qquad g\doteq \text{e}^{-a\Omega}.
\end{align}
Similarly, the covariance function is given by
\begin{align}
\E\left[(X_i-\E[X_i])(X_j-\E[X_j])\right]=\frac{b}{a}\left(g^{|i-j|}-g^{i+j}\right).
\end{align}
Therefore, for sufficiently large $i$, 
\begin{align}\LABEQ{OUmeanvar}
&X_i\sim \mathcal{N}(0,\frac{b}{a}),\\\LABEQ{OUcorr}
&\E\left[(X_i-\E[X_i])(X_j-\E[X_j])\right]=\frac{b}{a}\left(g^{|i-j|}\right).
\end{align}

If we recall the properties observed for the process $r_1(\tau)$ during the steady state phase:
\begin{align}
&\E[r_1(\tau)]=\hat{r_1}(*)=\gamma (\epsilon_{(l,r,L)}-\epsilon),\\
&\text{Var}[r_1(\tau)])=\frac{\delta_1^*}{M},\\
&\E[r_1(\zeta)r_1(\tau)]-\E[r_1(\zeta)]\E[r_{1}(\tau)]=\frac{\delta_1^*}{M} \text{e}^{-\theta|\tau-\zeta|},
\end{align}
we can conclude that the process $r_1(\tau)-\hat{r}_1(*)$ can
be identified as an OU process with parameters:
%
%
%
%The OU process presents constant mean and variance and an exponentially
%covariance decay and, thus, it constitutes a consistent model to
%the evolution of $r_1(\tau)$ in the steady state phase of the PD process.  By using  $\E[r_1(\tau)]\approx\gamma(l,r)\Delta_\epsilon$, $\text{Var}(r_1(\tau))\approx \nu(l,r)M^{-1}$ and that $\text{CoVar}[r_1(\tau),r_1(\zeta)]\approx\nu M^{-1}\exp(\theta(l,r)|\tau-\zeta|)$, 
%the process $r_1(\tau)-\hat{r}_1(*)$ can
%be identified to as an OU process with parameters:
\begin{align}\LABEQ{r1param}
a=\theta, \quad 
b=\frac{\delta_1^*\;\theta}{M},
\end{align}  
where we have taken $\Omega=M^{-1}$, i.e., the normalized time for
a single PD iteration.

\subsection{First-passage time distribution}\LABSEC{FPT}

The  statistical  distribution of the first-passage time (FPT)  of
an Ornstein-Uhlenbeck process, i.e., the first time at which an OU
process is above a certain boundary $s$, is of interest in a variety
of fields \cite{OUBio1,OUFinance1}.  Mathematically, if $X(t)$ is
an OU process, the FPT for a  boundary $s$ is defined as
\begin{align}
T_s=\inf_{t\geq0}\{t: X(t)\geq s\}.
\end{align}

Unlike the case of a Brownian motion, analytic expressions for the
p.f.d. of $T_s$, $p_{T_s}(t)$, known to date are quite involved
and for specific applications there is a need to perform numerical
computations of the density \cite{OU-4}. As summarized by Alili,
Patie and Pedersen in \cite{Patie05}, three representations  have
been proposed for the first-passage time density of an OU-process
through a constant boundary. The first expression is a series
expansion involving the eigenvalues of a Sturm-Liouville boundary
value problem associated with the Laplace transform of the FPT probability density function. The second one is an integral representation using its Fourier
transform, and the third one is given in terms of a functional of
a three-dimensional Bessel bridge. Also, expressions for the moments
of  $T_s$ are only known in integral form \cite{OU-1,OU-2}.

Assuming $X(\tau)=r_1(\tau)-\hat{r}_1(*)$ is a zero-mean OU process during the steady state phase, of length $\epsilon L-\tau^*$,  to estimate the error probability we have  to compute the cumulative probability
\begin{align}\LABEQ{PT}
P(0\leq T_s\leq \epsilon L-\tau^*)=\int_{0}^{\epsilon L-\tau^*}p_{T_s}(t)~ dt
\end{align}
for $s=\hat{r}_1(*)=\gamma(\epsilon_{(l,r,L)}-\epsilon)$. An estimate to \EQ{PT} can be obtained by numerical integration using the analytic expressions of $p_{T_s}(t)$ commented above. While in principle this approach is valid, we are rather interested in a more informative expression that provides insights into the relation between the SC-LDPC  structural parameters and the block error probability.

\begin{figure*}[!b]
\centering
\begin{tabular}{cc}
\includegraphics[scale=0.45]{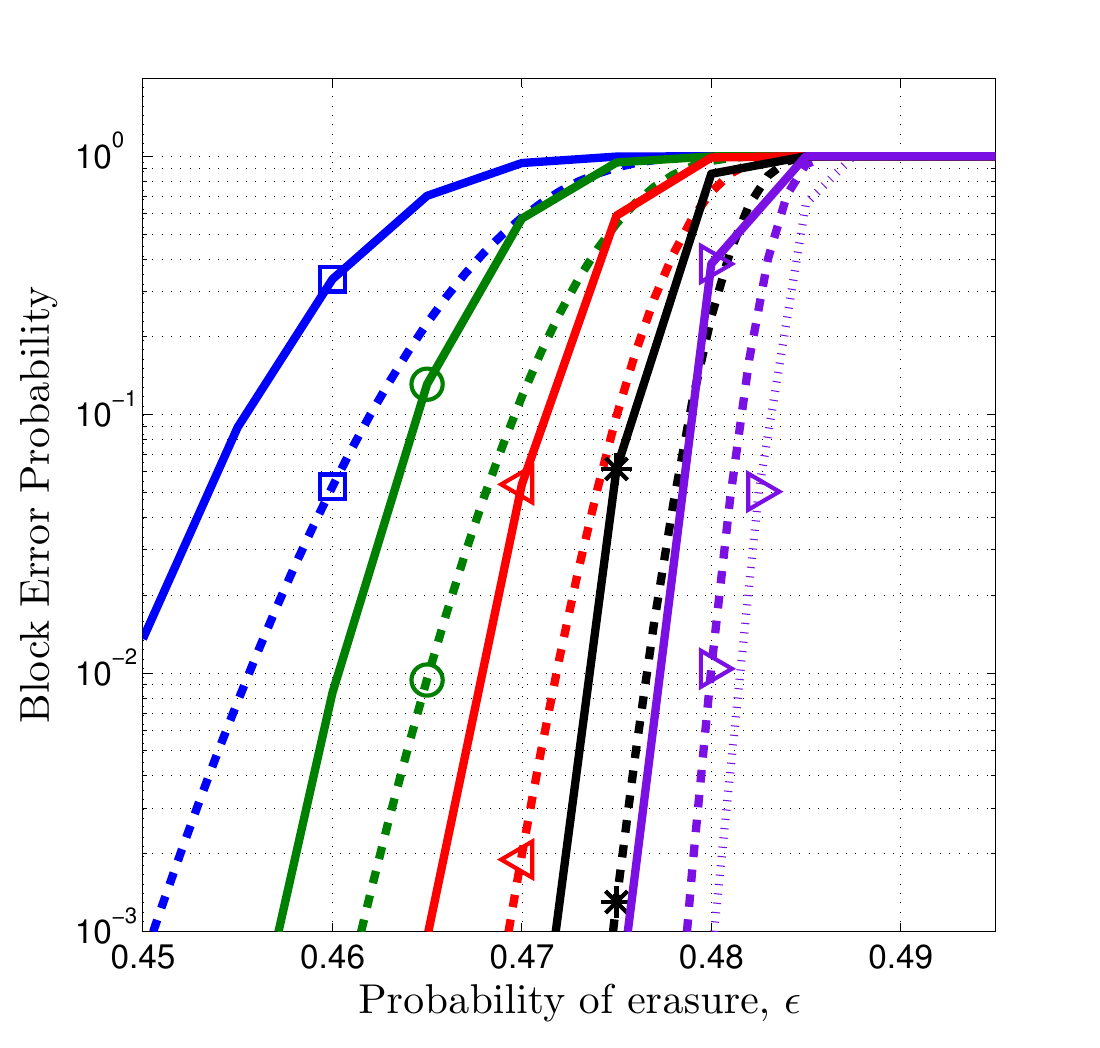}  & \includegraphics[scale=0.45]{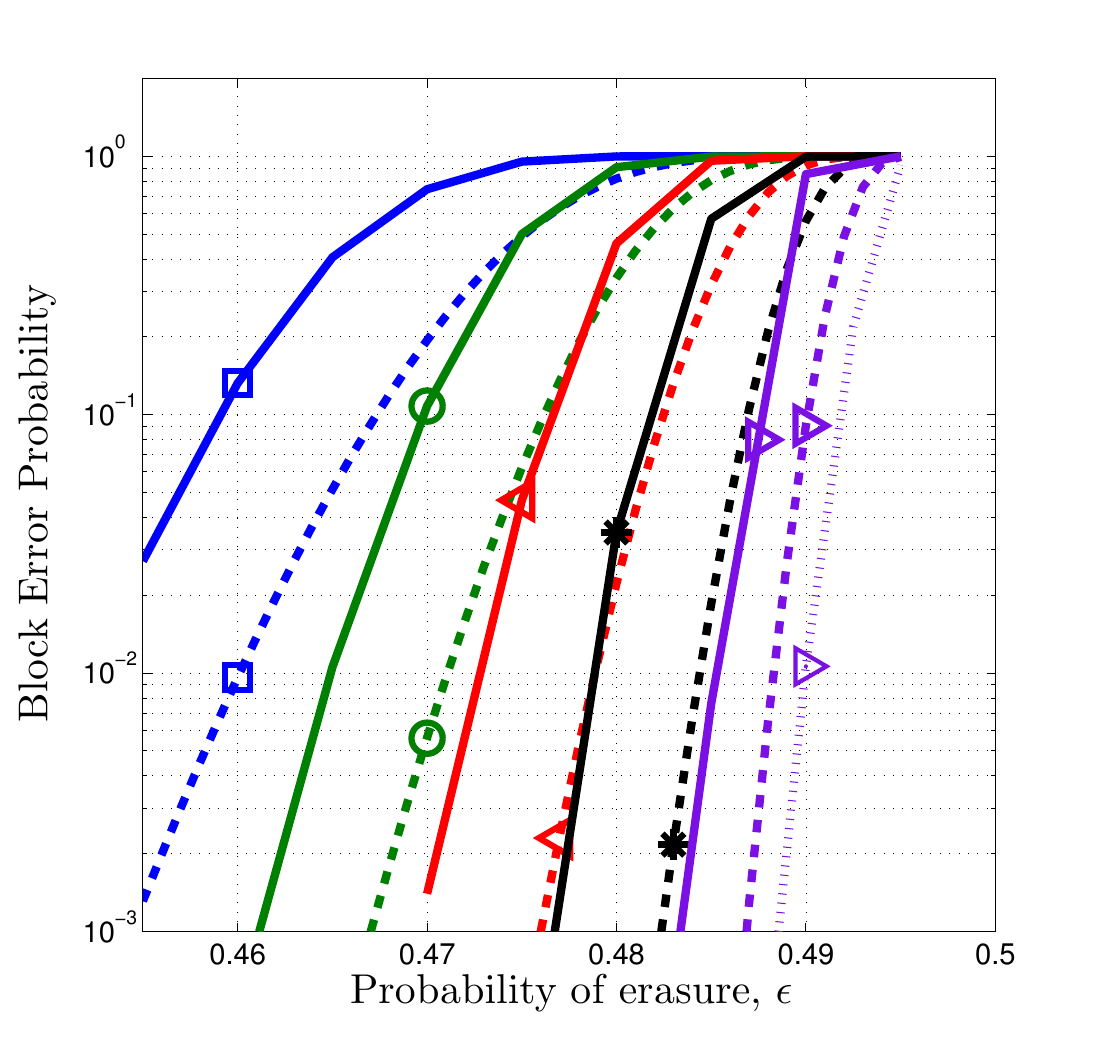}  \\
(a) & (b)
\end{tabular}
\caption{ Simulation error probability curves (solid lines) along
with the estimated performance using the expression in \EQ{SL} (dashed lines)
for the ensemble $(3,6,50)$ (a) and $(4,8,50)$ (b) with  $M=500$ $(\square)$, $M=1000$
$(\circ)$, $M=2000$ $(\lhd)$, $M=4000$ $(\ast)$ and $M=8000$ $(\rhd)$. For $M=8000$, we also include in dotted
lines the estimated performance  using the SL in \EQ{SL} along with
the upper bound to $\mu_0$ in \EQ{uppermu}.}\LABFIG{curves}
\end{figure*}

In \cite{OU-3}, the authors show that $p_{T_s}(t)$ converges as $\frac{s}{b/a}\rightarrow\infty$ to an exponential distribution
\begin{align}\LABEQ{PDF}
p_{T_s}(t)\sim\frac{1}{\mu_0}\text{e}^{-t/\mu_0},
\end{align}  
where $\mu_0=\E[T_s]$ is the OU mean first-passage time from the zero initial state  to the boundary $s$. In terms of the parameters of the $r_1(\tau)$ process in \EQ{r1param} we have
\begin{align}
\frac{s}{b/a}=\frac{\gamma M (\epsilon_{(l,r,L)}-\epsilon)}{\delta_1^*}\leq \frac{\gamma M \epsilon_{(l,r,L)}}{\delta_1^*}.
\end{align}
and hence the exponential distribution in \EQ{PDF} is achieved in the limit $M\rightarrow\infty$. In addition, $\mu_0$ can be exactly computed using the following integral expression \cite{OU-1}:
\begin{align}\LABEQ{mu0}
\displaystyle\mu_0(a,b,s)=\frac{\sqrt{2\pi}}{a}\displaystyle\int_{0}^{\frac{s}{\sqrt{\frac{b}{a}}}}\Phi(z)\text{e}^{\frac{1}{2}z^2}\text{d}z,
\end{align}
where $\Phi(z)$ is the c.d.f. of a Gaussian distribution with zero mean and unit variance, i.e.,
\begin{align}\label{equ:cdf}
\Phi(z) = \int_{-\infty}^z \frac{1}{\sqrt{2 \pi}} e^{-\frac{x^2}{2}} dx.
\end{align}
In the limit $s\rightarrow\infty$ (the same limit for which the
exponential distribution in \EQ{PDF} holds), the following upper
bound to \EQ{mu0} becomes tight:
\begin{align}\LABEQ{bound}
\mu_0(a,b,s)< \frac{\sqrt{2\pi}}{\sqrt{ab}} s ~\exp\left(\frac{s^2}{2\frac{b}{a}}\right)
\end{align}

\proof{Denote $C=\frac{s}{\sqrt{\frac{b}{a}}}$. Since $\Phi(z)\in[0,1]$, 
\begin{align}\nonumber
\mu_0(a,b,s)&=\frac{\sqrt{2\pi}}{a}\int_{0}^{C}\Phi(z)\text{e}^{\frac{1}{2}z^2}\text{d}z\\ &< \frac{\sqrt{2\pi}}{\theta} \int_{0}^{C}\text{e}^{\frac{1}{2}z^2}\text{d}z\stackrel{(a)}{=}\frac{\sqrt{2\pi}}{a} \sum_{n=0}^{\infty}\frac{2^{-n}~C^{2n+1}}{n!(2n+1)},\nonumber\\ 
&=\frac{\sqrt{2\pi}}{a} C \sum_{n=0}^{\infty} \frac{(C^2/2)^n}{n!(2n+1)}\LABEQ{limalternative2}
\end{align}
where the equality $(a)$ is obtained by using the series expansion $\exp(x)=\sum_n x^{n}/n!$ Further,  \EQ{limalternative2} can be  upper bounded as follows:
\begin{align}
&\frac{\sqrt{2\pi}}{a} C \sum_{n=0}^{\infty} \frac{(C^2/2)^n}{n!(2n+1)}\nonumber\\
&<\frac{\sqrt{2\pi}}{a} C \sum_{n=0}^{\infty} \frac{(C^2/2)^n}{n!}\nonumber\\
&=\frac{\sqrt{2\pi}}{a} C \exp(C^2/2)=\frac{\sqrt{2\pi}}{\sqrt{ab}} s ~\exp\left(\frac{s^2}{2\frac{b}{a}}\right).
\end{align}
\begin{align*}
\qquad \qquad \qquad \qquad \qquad \qquad \qquad \qquad \qquad \qquad \qquad \qquad \blacksquare
\end{align*}
}
%
%
%
%
%
%
% statistical moments of $T_s$. In principle the p.d.f. of $T_s$ can be reconstructed via its moments \cite{OU-4}. However, for a given zero-mean OU process $X(t)$ with variance $b/a$, as $\frac{s}{b/a}\rightarrow 0$, even the numerical evaluation of the c.d.f. of $T_s$  becomes a complex problem. In our context,
%$s$ is given by $\hat{r}_1(*)=\gamma\Delta_\pe$ and
%\begin{align}
%\frac{s}{b/a}=\frac{\gamma\Delta_\pe M}{\delta_1^*}.
%\end{align}Section\SEC{FPT},
%Hence, $\frac{s}{b/a}\rightarrow0$ represents the case in which either
%$M$ is too small or we are too close to the BP threshold. In both cases we know the error probability is close to one. In \cite{OU-3},
%the authors show that the first-passage time p.d.f. $p(T_s)$ converges with growing
%$\frac{s}{b/a}$  to an exponential distribution
%\begin{align}\LABEQ{PDF}
%p(T_s)\sim\frac{1}{\mu_0}\text{e}^{-t/\mu_0},
%\end{align}  
%where $\mu_0$ is the OU mean first-passage time from the zero initial state  to the boundary $s$. An explicit expression is known for $\mu_0$ 
%\cite{OU-1,OU-3}:
%\begin{align}
%\displaystyle\mu_0=\frac{\sqrt{2\pi}}{a}\displaystyle\int_{0}^{\frac{s}{\sqrt{\frac{b}{a}}}}\Phi(z)\text{e}^{\frac{1}{2}z^2}\text{d}z,
%\end{align}
%where $\Phi(z)$ is the c.d.f. of a Gaussian distribution with zero mean and unit
%variance. 

In our decoding scenario, we model the process $r_1(\tau)-\hat{r}_1(\tau^*)$ as an OU process with parameters given in \EQ{r1param}. Hence, $\mu_0$ represents the average survival  time once $r_1(\tau)$ has entered the steady state phase. Using \EQ{r1param}, we obtain
\begin{align}\LABEQ{tauerr}
\mu_0=\frac{\sqrt{2\pi}}{\theta}\int_{0}^{\frac{\gamma}{\sqrt{\delta^*_{1}}}\sqrt{M}\Delta_{\epsilon}}\Phi(z)\text{e}^{\frac{1}{2}z^2}\text{d}z,
\end{align}
where  $\Delta_{\epsilon}=(\epsilon_{(l,r,L)}-\epsilon)$. Note that the integral in  \EQ{tauerr} diverges; $\mu_0$, the expected time at which the process $r_1(\tau)$ dies, grows exponentially fast with $M$ and $\Delta_{\epsilon}$. By \EQ{bound}, for large $M\Delta_{\epsilon}$, $\mu_0$ is tightly upper bounded by
\begin{align}\LABEQ{uppermu}
\mu_0<\frac{\sqrt{2\pi}}{\theta}\frac{\gamma}{\sqrt{\delta^*_{1}}} \sqrt{M}\Delta_{\epsilon}\exp\left(M\frac{\gamma^2\Delta_{\epsilon}^2}{2\delta_1^*}\right).
\end{align}

By taking a sufficiently large $M$, we can make $\mu_0>>\epsilon
L$, which means that the length of the critical phase is very short
compared to the time that, in average, we have to wait until
$r_1(\tau)$ takes zero value. This  results in a small error
probability.

%As shown in Section \SEC{CovTime}, we can design the SC-LDPC coupling pattern to make $\theta$ smaller and, therefore, to increase $\mu_0$. However, note that the sensibility of $\mu_0$ w.r.t. $\theta$ is not as strong as its sensibility w.r.t. to $M$. Also, while designing the SC-LDPC coupling pattern to obtain smaller $\theta$ while keeping the coding rate unaltered seems like a hard problem, 
%
%
%the sensibility of $\mu_0$ w.r.t. to $\theta$ is 

%\begin{figure}[h]
%\centering \includegraphics[scale=0.5]{probs_36.eps}
%\caption{ Simulation error probability curves (solid lines) along
%with the prediction using the expression in \EQ{SL} (dashed lines)
%for the ensemble $(3,6,50,M)$ and $M=500$ $(\ast)$, $M=1000$
%$(\circ)$, $M=2000$ $(\rhd)$ and $M=4000$ $(\times)$.}\LABFIG{curves}
%\end{figure}

\subsection{\textcolor{black}{Scaling law for the $(l,r,L)$ ensemble}}

The first-passage time probability distribution in \EQ{PDF} along
with \EQ{tauerr} constitutes the necessary tools to estimate the
decoding error probability. The steady state phase roughly lasts
between $\tau^*=\tau^*(l,r,L,\epsilon)$ and $\epsilon L$.  Recall
also that $\tau^* $ is lower bounded by $\underline{\tau}$, computed
in Section\SEC{lower} from the DE solution of the uncoupled
$(l,r)$-regular LDPC code ensemble. We estimate the $(l,r,L)$
ensemble average probability using the exponential distribution
c.d.f., obtaining the following \emph{scaling law} (SL):
\begin{align}\LABEQ{SL}
&\E_{(l,r,L)}[P_{B}(l,r,L,M,\epsilon)]\nonumber\\
&\approx 1-\exp\left(-\frac{\epsilon L- \tau^*}{\mu_0}\right)\nonumber\\
&\leq 1-\exp\left(-\frac{\epsilon
L- \underline{\tau}}{\mu_0}\right)\nonumber\\
&=1-\exp\left(-\frac{\epsilon L- \underline{\tau}}{\displaystyle \frac{\sqrt{2\pi}}{\theta}\int_{0}^{\frac{\gamma}{\sqrt{\delta^*_{1}}}\sqrt{M}\Delta_{\epsilon}}\Phi(z)\text{e}^{\frac{1}{2}z^2}\text{d}z}\right),
\end{align} where,
$\E_{(l,r,L)}[P_{B}(l,r,L,M,\epsilon)]$ represents the expected
block error probability of the $(l,r,L)$ ensemble with $M$
bits per position averaged over all codes in the ensemble and all
channel realizations and $\Phi(z)$ is given in (\ref{equ:cdf}).

Note that the expression in \EQ{SL} depends on the ensemble parameters $L$ and $M$ and
on the following \emph{scaling parameters}:
\begin{enumerate}
\item gap to threshold $\Delta_{\pe}=\pe_{(l,r,L)}-\pe$
\item mean parameter $\gamma$ (see Table~\TAB{TABLAGAMMA})
\item variance parameter $\delta_1^*$ (see Table~\TAB{TABLANU})
\item rate of correlation decay $\theta$ (see Table~\TAB{thetas})
\end{enumerate}
In addition, there is a (slight) dependence on $\underline{\tau}$,
see Table\TAB{TABLAtau}.

\vspace{0.1cm}
In Fig.~\FIG{curves} we compare simulated block error probability
curves (solid lines) along with the prediction using the expression
in \EQ{SL} (dashed lines) for the ensembles $(3,6,50)$  (a) and
$(4,8,50)$ (b). The number $M$ of bits per position is: $M=500$
$(\square)$, $M=1000$ $(\circ)$, $M=2000$ $(\lhd)$, $M=4000$ $(\ast)$
and $M=8000$ $(\rhd)$.  For $M=8000$, we also include in dotted
lines the estimated performance  using the SL in \EQ{SL} along with
the upper bound to $\mu_0$ in \EQ{uppermu}.

First of all, we can see that there is a systematic ``shift'' between
the actual error rate curves and the curves predicted via our scaling
law. The estimate in  \EQ{SL} heavily relies on the assumption that
the p.d.f. of the first-passage time of the $r_1(\tau)$ is distributed
according to an exponential distribution. But this is  an asymptotic
result that only holds  in the limit $M\rightarrow\infty$. In order
to improve the estimate in \EQ{SL}, it is an interesting and
challenging problem to drop the exponential distribution assumption
and consider existing approaches to the p.d.f. of the first-passage
time of an OU process for finite values of the  boundary $s$
(controlled by $M$ in our problem) \cite{Patie05}. As we can observe
in Fig.~\FIG{curves}, for a few thousand bits per position, \EQ{SL}
provides an accurate estimate of the $(l,r,L)$ block error rate.
At these lengths, SC-LDPC codes are prominent candidates  for future
communication standards such as optical communications \cite{ChangFEC}
and wireless digital broadcasting \cite{DVBT2}.   Also, note that
in all cases the slope of the error rate curves computed using
\EQ{SL} matches with the slope of the simulated error rates for the
same $M$.

Hence, our first conclusion is that the $(l,r,L)$ performance can be
accurately estimated if $M$ is sufficiently large.
But our aim  is to also show that the SL in \EQ{SL}
captures the right scaling between the ensemble block error rate
and the different structural  parameters, in particular  the gap
to threshold, $M$ and $L$. In this regard, the SL can be used to
predict the performance improvement/degradation  when a 
parameter is modified around a certain value. For instance, assume
that for fixed $M$ and $L$, the waterfall performance of
the $(l,r,L)$ ensemble has been estimated via Monte
Carlo simulation. Now we can use the SL in \EQ{SL}  to predict the
performance variation when the chain length and/or the number of bits
per position are set to $L'=c L$ and $M'=k M$ respectively, $c,k\in\mathbb{R}^+$.
These type of calculations are  relevant from the practical point
of view and the scaling law in \EQ{SL} is a useful tool to perform
quick estimates.

\begin{figure}[t!]
\centering \includegraphics[scale=0.45]{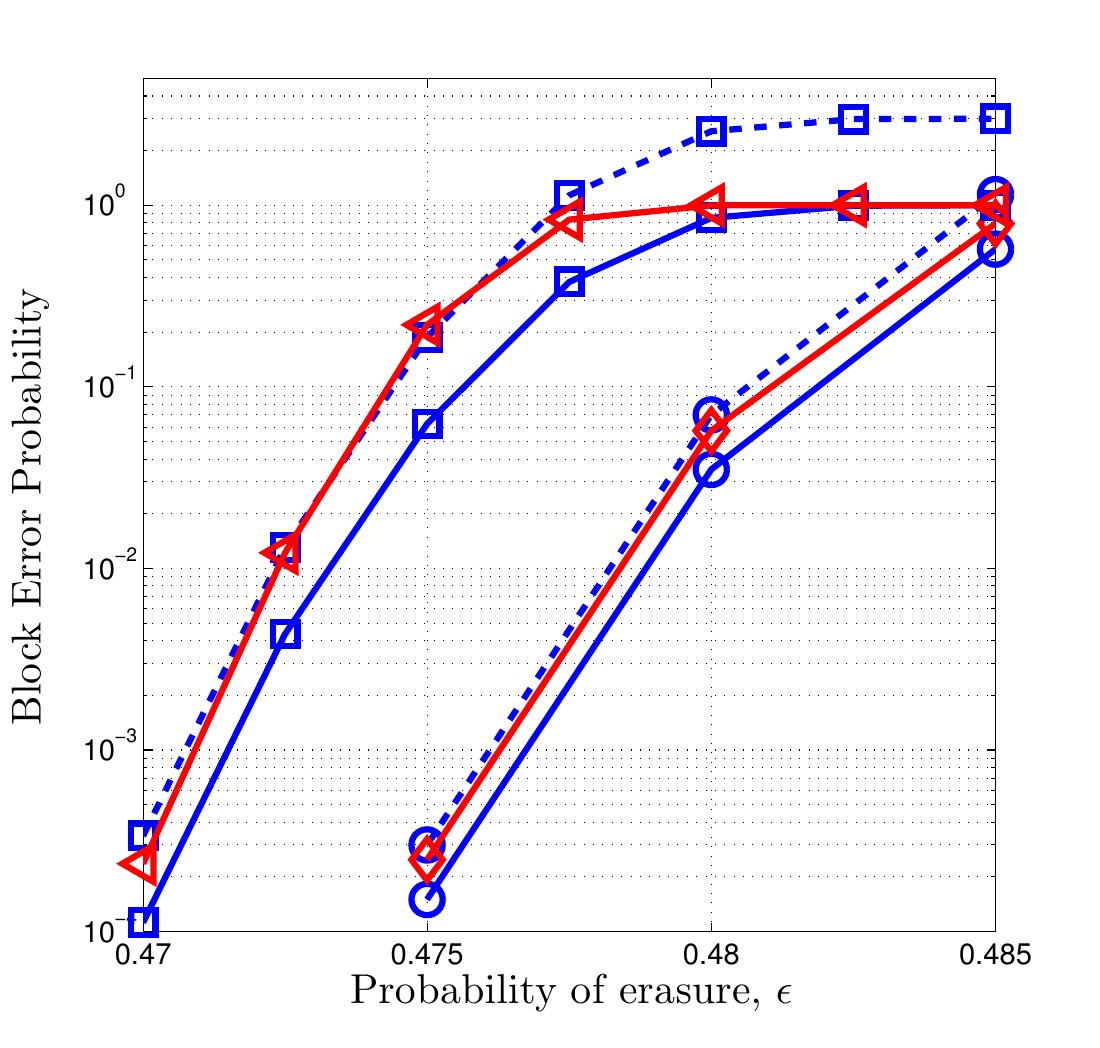}
\caption{ Simulated error probability curves (solid lines) for the ensembles $(3,6,50)$ ($\square$), $(3,6,150)$ ($\lhd$), $(4,8,50)$ ($\circ$) and $(4,8,100)$ ($\diamond$) with $M=4000$ bits. In dashed lines, we represent the estimated error probability for the ensembles $(3,6,150)$ and  $(4,8,100)$ computed by multiplying the simulated error rates for the ensembles $(3,6,50)$ and $(4,8,50)$ by three and two respectively.}\LABFIG{curves2}
\end{figure}

\begin{figure*}[h!]
\centering
\begin{tabular}{cc}
\includegraphics[scale=0.45]{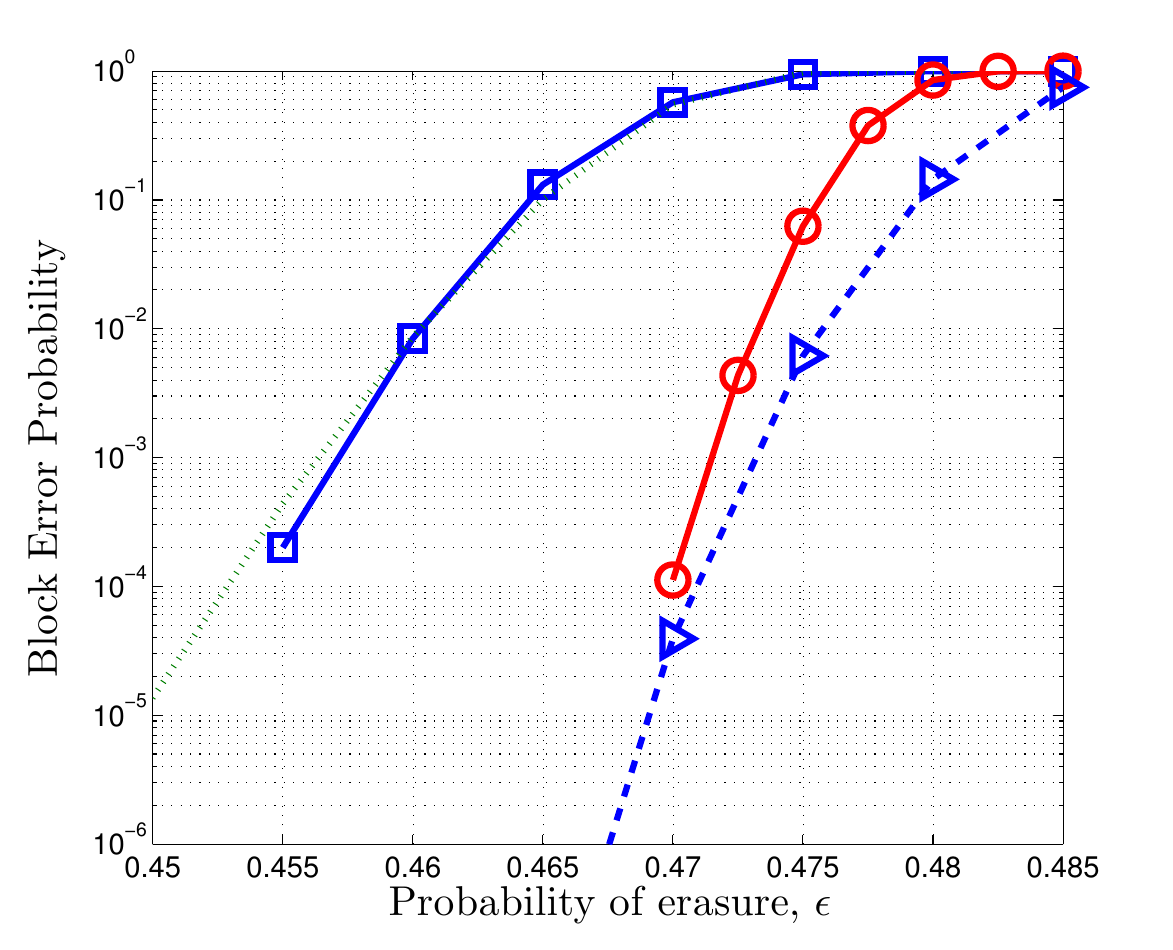}  & \includegraphics[scale=0.45]{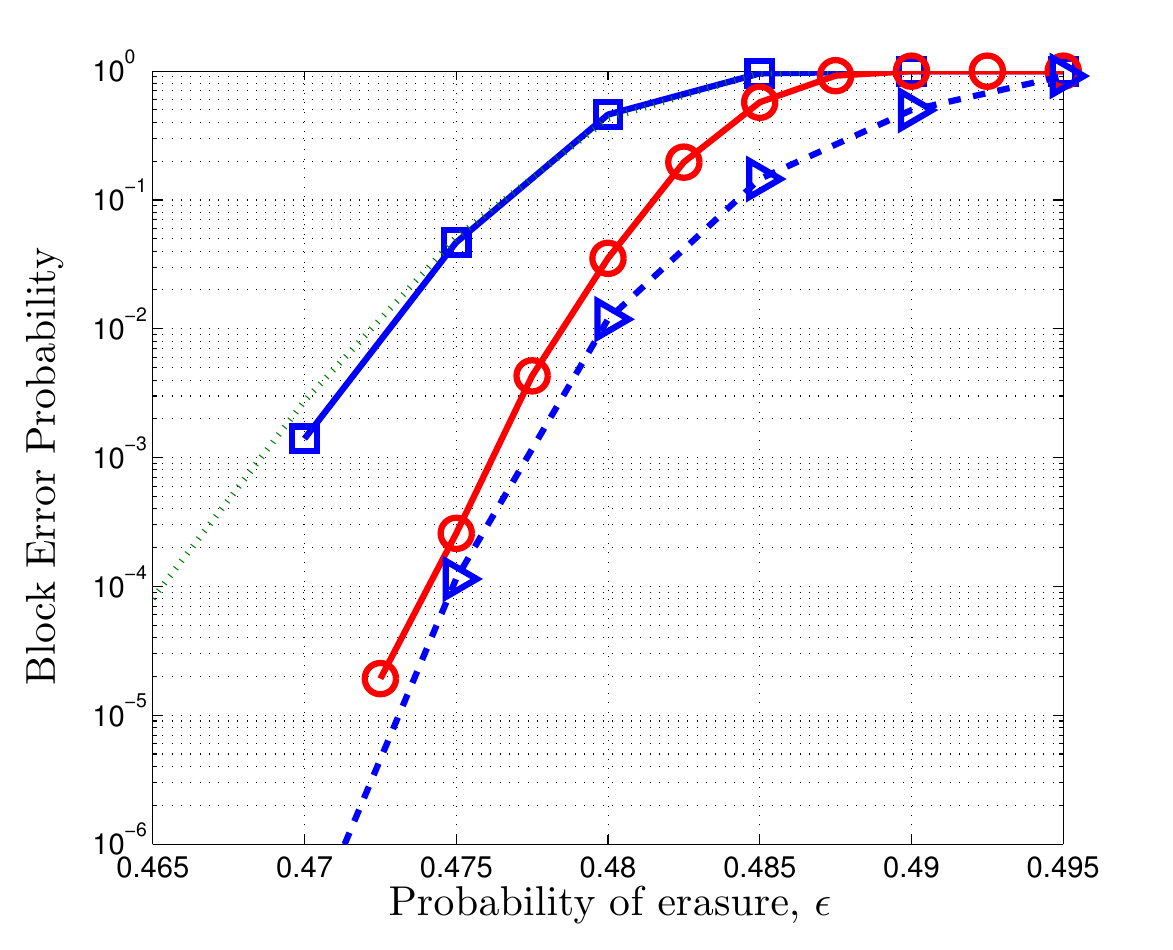}  \\
(a) & (b)
\end{tabular}
\caption{ In Fig. \FIG{curvesscalingM}, we represent the simulated performance for the $(3,6,50)$ ensemble with $M=1000$ bits $(\square)$ (a) and the  $(4,8,50)$ ensemble with $M=2000$ bits $(\square)$ (b). In both cases, the dashed lines with $(\rhd)$ marker represents the estimated performance using \EQ{SLlow2} for  $M=4000$ bits. The actual $M=4000$ performance computed by simulation for each case is given by the solid line with $(\circ)$ maker.}\LABFIG{curvesscalingM}
\end{figure*}
We will illustrate the problem in the low error rate regime. If we take a sufficiently large $M$,  the error probability predicted by \EQ{SL} is small and we can use a first order Taylor expansion to simplify the expression:
\begin{align}\LABEQ{SLlow}
\E_{(l,r,L)}[P_{B}(l,r,L,M,\epsilon)]\approx \frac{\epsilon L- \underline{\tau}}{\mu_0}.
\end{align}
In addition, in this regime the bound to $\mu_0$ in \EQ{uppermu} is tight and hence, 
\begin{align}\LABEQ{SLlow2}
\E_{(l,r,L)}[P_{B}(l,r,L,M,\epsilon)]&\approx \frac{ \theta \sqrt{\delta_1^*} (\epsilon L- \underline{\tau})}{\sqrt{2\pi}\gamma \sqrt{M}\Delta_{\epsilon}} \exp\left(-M\frac{\gamma^2 \Delta_{\epsilon}^2}{2 \delta_1^*}\right).
\end{align}
Note that this expression indicates that the error probability scales linearly with the chain length. In Fig. \FIG{curves2}, we represent the simulated  error probability curves (solid lines) for the ensembles $(3,6,50)$ ($\square$), $(3,6,150)$ ($\lhd$), $(4,8,50)$ ($\circ$), $(4,8,100)$ ($\diamond$) with $M=4000$ bits. In dashed lines, we represent the estimated error probability for the ensembles $(3,6,150)$ and  $(4,8,100)$ computed by multiplying the simulated error rates for the ensembles $(3,6,50)$ and $(4,8,50)$ by three and two respectively. 
Observe that, when the error probability is small enough the predicted performance matches with the simulated one for $L=150$ and  $L=100$, indicating that the scaling in \EQ{SLlow} w.r.t $L$ is essentially correct. 

Now we test whether the scaling with respect to $M$ predicted by  \EQ{SLlow2} is also accurate. Consider the $(l,r,L)$ ensemble with $M$ bits per position. Assume that we numerically find  $M_{\text{SL}}<M$ such that the simulated performance for $M$ bits per position and the one predicted by \EQ{SL} with $M_{\text{SL}}$ match.  Using the SL, we can now estimate how much the performance is improved when the number  of bits per position is modified from $M$ to $kM$, for some $k>0$. Using the simplified expression in \EQ{SLlow2} for the low error rate regime, a simple calculation shows that
\begin{align}\LABEQ{scalingM}
&\E_{(l,r,L)}[P_{B}(l,r,L,kM,\epsilon)]\\\nonumber\\\nonumber
&\approx \frac{\E_{(l,r,L)}[P_{B}(l,r,L,M,\epsilon)]}{\sqrt{k}}\exp\left(-M_{\text{SL}}\frac{\gamma^2 (k-1) \Delta_{\epsilon}^2}{2\delta_1^*}\right).
\end{align}

The accuracy of the estimate in \EQ{scalingM} in the low error rate regime is demonstrated in Fig. \FIG{curvesscalingM}. In Fig. \FIG{curvesscalingM}(a), we represent the simulated performance for the $(3,6,50)$ ensemble with $M=1000$ bits $(\square)$, solid line. The curve obtained is well approximated by \EQ{SL} with $M_{\text{SL}}=700$ bits (dotted line). Using this value, the dashed line with $(\rhd)$ marker represents the estimated performance  for the $(3,6,50)$ ensemble with $M=4000$ bits using \EQ{SLlow2}, whereas the actual simulated performance for this case is plotted with solid line and $(\circ)$ marker. As we can observe, the prediction is indeed accurate for small error rates. In Fig. \FIG{curvesscalingM}(b), we show a similar example using the $(4,8,50)$ ensemble. The solid curve with $(\square)$ marker represents the simulated performance for the case $M=2000$ bits. This curve is well approximated  by \EQ{SL} with $M_{\text{SL}}=1100$ bits and this value is used to compute the estimated performance for $M=4000$ bits, $(\rhd)$ maker with dashed line. The actual $(4,8,50)$ performance is plotted with $(\circ)$ maker. In all cases, the simulated performance curves have been obtained after $10^5$ transmitted codewords. The values for $\gamma$ and $\delta_1^*$ in  \EQ{SLlow2} are given for each ensemble in Table\TAB{TABLAGAMMA} and Table\TAB{TABLANU}.

The above results show that the SL proposed is consistent with the scaling behavior of the code. In \cite{Olmos11-3}, the performance of a class of SC-LDPC codes constructed from protographs  was tested for a wide range of scaling functions $L=f(M)$. It was observed that the ensemble presented a vanishing error probability in the limit $M\rightarrow\infty$  for $\pe$ values very close to the MAP threshold of the uncoupled ensemble even though when $L$ grows much faster than $M$. In particular, we conjectured that this effect is lost only when the  chain length $L$ grows at least  exponentially fast with $M$. The scaling law for the $(l,r,L)$ ensemble in \EQ{SL} is consistent with such a conjecture. Indeed, if $L=f(M)$ and $M$ is large enough then the SL  in \EQ{SL} takes the form:
\begin{align}\LABEQ{SLlow3}
 1-\exp\left(-\frac{\epsilon f(M)-\tau^*}{\frac{\sqrt{2\pi}}{\theta}\frac{\gamma}{\sqrt{\delta^*_{1}}} \sqrt{M}\Delta_{\epsilon}\exp\left(\frac{\gamma^2 M\Delta_{\epsilon}^2}{2\delta_1^*}\right)}\right)
\end{align}
and, hence, we can easily find a pair of real positive constants for which a scaling of the form $L=a\exp(bM)$ makes \EQ{SLlow3} tend to one in the limit $M\rightarrow\infty$. This result indicates that the wave-like decoding  phenomenon, which allows achieving near-capacity thresholds, is very robust.

\subsection{\textcolor{black}{Finite-length performance and scaling parameters}}

Four  scaling parameters appear in the scaling law proposed
in \EQ{SL}, as well as in the low-error rate approximation in \EQ{SLlow2}.
In this regard, let us conclude this paper by briefly discussing
the sensitivity of the SC-LDPC finite-length performance with respect
to each one of them:
\begin{enumerate}
\item Mean  parameter $\gamma$ and variance parameter $\delta_1^*$. First of all, note that these two parameters only appear in the SL by means of the ratio $\alpha\doteq\gamma/\sqrt{\delta_1^*}$. By \EQ{SLlow2}, note the performance is very sensitive w.r.t. to any change in $\alpha$. A simple calculation shows that
\begin{align}
&\frac{\partial \E_{(l,r,L)}[P_{B}(l,r,L,M,\epsilon)]}{\partial \alpha}\sim \mathcal{O}\left(-\text{e}^{-\alpha^2 \Delta_{\epsilon} M}\right),
\end{align}
and thus a small increase in $\alpha$ can lead to a significant
decay in the error rate. On one hand, we have shown how $\gamma$ is directly proportional to the speed of the wave under BP message-passing decoding. From the design perspective, it has been shown how this speed can be improved by  optimizing the ensemble degree profile \cite{Vahid2013}. On the other hand, the design of the SC-LDPC code ensemble to reduce the variance parameter $\delta_1^*$ is still a challenging open
question since no closed-form expressions to compute $\delta_1^*$ (or bound
it) have been proposed yet.

\item SC-LDPC code ensemble threshold.  As in the case of $\alpha$, the
finite-length performance is quite sensitive to $\Delta_{\epsilon}$.
Since it is fairly easy to design  SC-LDPC code ensembles with
capacity-approaching thresholds \cite{Lentmaier10,Lentmaier10-2},
the main question here is to determine whether the optimization of
the SC-LDPC code ensemble to achieve thresholds arbitrarily close to
capacity is compatible with the optimization of the rest of parameters
that appear in the scaling law.

\item $\theta$ parameter. As described in Section\SEC{CovTime}, $\theta$ is related to the covariance decay of the $r_1(\tau)$ process along the time and can be controlled by the coupling pattern. Intuitively, the more we spread the connections of variable nodes along different positions, the smaller $\theta$ value we can observe. However, it is not a trivial problem how to optimize this parameter at fixed coding rate (recall the coupling pattern also determines the boundary conditions, and thus, the coding rate loss incurred). In addition, upon optimization we might not observe a significant improvement in the finite-length performance since by \EQ{SLlow2} it only scales linearly with $\theta$.
\end{enumerate}

\section{Conclusions and Future Work}\LABSEC{Conclusions}

We have analyzed the decoding of finite-length SC-LDPC codes over
the BEC using the peeling decoder. By extending the methodology
applied in \cite{Urbanke09} to study finite-length LDPC code ensembles,
we have determined and solved covariance evolution, i.e., the system
of differential equations that encodes the first two statistical
moments of the evolution of the residual graph. We have shown that
the statistical evolution of the normalized number  of degree-one
check nodes along the peeling decoding process, i.e., the $r_1(\tau)$
process, is of a completely different nature compared to the case
of the uncoupled $(l,r,)$-regular LDPC code ensemble.  In the coupled
ensemble, decoding failures more or less happen uniformly throughout
the whole decoding process (with the exception of the very beginning
and the very end of the process).  Furthermore, the survival
probability of the process $r_1(\tau)$ can be estimated by assuming
that the $r_1(\tau)$-process is in fact a properly chosen
Ornstein-Uhlenbeck process. The scaling law derived via this
assumption contains four parameters: the ensemble threshold, the
mean parameter $\gamma$, the covariance parameter $\delta_1^*$ and
the correlation parameter $\theta$.  The threshold can be determined
through standard DE techniques.  We derived the remaining parameters,
namely $(\gamma,\delta_1^*,\theta)$, via covariance evolution.

The closed-form SL proposed in this paper provides code designers
with some insight into on the main scaling behavior between the
error rate and the different code parameters.  The scaling law gives
a good approximation to the actual performance (determined via
simulations) and, furthermore, it can be used to accurately predict
how the performance changes if we modify some base parameters, such
as $L$ or $M$.

Several important practical questions remain open.  First of all,
it would be interesting to determine whether there exist alternative
methods to evaluate the triple $(\gamma,\delta_1^*,\theta)$ efficiently.
As we mentioned in the paper, at least the parameter $\gamma$ is
directly related to the decoding speed, and this decoding speed in
term can be computed via basic quantities that appear in density
evolution. If it was able to derived closed form expressions for these parameters
then in principle this would open the door to perform an optimization
of the code parameters.  Also, what are the trade-offs between the
parameters $(\gamma,\delta_1^*,\theta)$? As we have discussed, we would
like $\gamma$ to be large and $\theta$ and $\delta_1^*$ to be small. But most likely, these are conflicting goals.

Finally, we have ignored some relevant aspects. Most importantly we
ignored the effect of the termination on the error probability, the
effect of a windowed-decoding algorithm \cite{Iyengar11}, or the
effect of the exact code structure.  All of these are likely going
to play a role in practical applications and hence have to be
considered.  In particular, we should address the analysis of
protograph-based SC-LDPC codes \cite{Lentmaier10, Stinner14} and
spatially-coupled structures based on generalized LDPC codes
\cite{Mitchell13GLDPC} or non-binary LDPC codes \cite{Huang14}.

\bibliography{allbib20_03.bib}
\bibliographystyle{IEEEtran}
% that's all folks

\appendices
\section{Expected evolution in one iteration of the PD}\LABSEC{A1}
Here we show how to compute the expected
evolution of the $(l,r,L)$ residual graph DD in one PD iteration,  namely
\begin{align}
&\E[R_{j,u}(\tau+1/M)-R_{j,u}(\tau)|\G(\tau)],\\
&\E[V_u(\tau+1/M)-V_{u}(\tau)|\G(\tau)],
\end{align}
for $u\in[D]$ and $j\in[r]$. To keep the notation uncluttered, we omit the conditionality on $\G(\tau)$ and we denote 
\begin{align}
\Delta R_{j,u}(\tau)&=R_{j,u}(\tau+1/M)- R_{j,u}(\tau),\\
\Delta V_{u}(\tau)&=V_{u}(\tau+1/M)-V_{u}(\tau).
\end{align}
Let $\mathtt{pos}(\tau)$ be the position at which we remove a degree-one check node at
time $\tau$.  The actual
check node that is removed is chosen uniformly at random from all
degree-one check nodes at this position.  The resulting probability
distribution $P(\mathtt{pos}(\tau)=u)$ is described by $p_u$ in
\EQ{pu}. 

Assume that $\mathtt{pos}(\tau)=m$.  Given the properties of the $(l,r,L)$ ensemble described in Section\SEC{S1}, it follows that the
variable connected to this  degree-one check node is placed
at position $u$ with probability
\begin{align}
\lambda_{m,u}(\tau)=\frac{V_u(\tau)}{\sum_{i=m-(l-1)}^{m}V_i(\tau)}
\end{align}
if $u\in\{m-(l-1),\ldots,m\}$ and  zero otherwise.  By extension, when a degree-one check node
from position $m$ and the variable connected to it are removed, then with probability
\begin{align}\LABEQ{xi}
\xi_{m,u}(\tau)=\sum_{i=u-(l-1)}^{u}\lambda_{m,i}(\tau).
\end{align}
one of the $l$ removed edges is connected to a check node at position $u$, $u\in[D]$. Note that $\xi_{u,u}(\tau)=1$ $\forall u$.

Given these definitions we can compute the expected evolution of the DD conditioned to the case for which the degree-one check node is removed from position $m$, i.e., $\mathtt{pos}(\tau)=m$. At position $m$, the variation in the DD is simple: $\E[\Delta R_{j,m}(\tau)|\mathtt{pos}(\tau)=m]$ is equal to $-1$ for $j=1$ and zero otherwise.
%The expected DD evolution
%at position $m=\mathtt{pos}(\tau)$ is straightforward to compute
%since we just remove an edge connected to a check of degree one. Thus,
%$\E[R_{j,m}(\tau+1)-R_{j,m}(\tau)|\mathtt{pos}(\tau)=m]$ is equal
%to $-1$ for $j=1$ and zero otherwise. 
At any other position $u\neq m$, an edge is removed from position 
$u$ with probability
$\xi_{m,u}(\tau)$. The edge removed is connected to a check node
of degree $j$ with probability
\begin{align}
\displaystyle\frac{R_{j,u}(\tau)}{\sum_{q=1}^{r}R_{q,u}(\tau)},
\end{align}
and, if this happens, $j$ edges of right degree $j$ are removed from the graph at position $u$. Also, $j-1$ edges of right degree $j-1$ are created at the same position. Hence, the expected
graph evolution at position $u\neq m$  is 
\begin{align}
\E[\Delta R_{j,u}(\tau)|\mathtt{pos}(\tau)=m]=j\xi_{m,u}(\tau)\frac{R_{j+1,u}(\tau)-R_{j,u}(\tau)}{\sum_{q=1}^{r}R_{q,u}(\tau)},
\end{align}
where for $j=r$, $R_{j+1,u}(\tau)=0$. Under the same assumption, 
$\mathtt{pos}(\tau)=m$, a variable node is removed from position $u$ with
probability $\lambda_{m,u}(\tau)$. Therefore,
\begin{align}
&\E[\Delta V_u(\tau)|\mathtt{pos}(\tau)=m]=-\lambda_{m,u}(\tau),
\end{align}
and this holds for $u\in[D]$, including the case $u=m$.

Finally, the expected graph evolution after one iteration of the
PD can be computed as follows. For $2\leq j\leq (r-1)$ we get,
\begin{align}\LABEQ{ExRDD}
&\E[\Delta R_{j,u}(\tau)]\\
\nonumber
&=j\sum_{\substack{m=1\\m\neq u}}^{D} \xi_{m,u}(\tau)\frac{R_{j+1,u}(\tau)-R_{j,u}(\tau)}{\sum_{q=1}^{r}R_{q,u}(\tau)}p_m(\tau)\\\nonumber
&=j\left(\frac{R_{j+1,u}(\tau)-R_{j,u}(\tau)}{\sum_{q=1}^{r}R_{q,u}(\tau)}\right)\left(\boldsymbol{p}^{T}\boldsymbol{\xi}_{u}-p_u(\tau)\right),
\end{align}
where $\boldsymbol{p}\doteq\left[p_{1}(\tau) \ldots p_{D}(\tau)\right]$
and $\boldsymbol{\xi}_{u}=[\xi_{1,u},\xi_{2,u}, \ldots,\xi_{D,u}]$ are $D$-length vectors.
For the case $j=1$, we also have to take into account the degree-one check node removed from the graph, which belongs to position $u$ with probability $p_u(\tau)$. Thus, 
\begin{align}\LABEQ{f1}
&\E[\Delta R_{1,u}(\tau)]=\\\nonumber
&=-p_u(\tau)+\left(\frac{R_{2,u}(\tau)-R_{1,u}(\tau)}{\sum_{q=1}^{r}R_{q,u}(\tau)}\right)\left(\boldsymbol{p}^{T}\boldsymbol{\xi}_{u}-p_u(\tau)\right).
\end{align}
On the variable side we obtain 
%On the variable side we obtain:
\begin{align}\LABEQ{f2}
\E[\Delta V_u(\tau)]&=-\sum_{m=1}^{D} \lambda_{m,u}(\tau)p_m(\tau).
\end{align}

\section{Covariance Evolution in one iteration of the PD}\LABSEC{A2}

We are now interested in computing the second order moments for the $(l,r,L)$ ensemble residual DD transition at one PD iteration:
\begin{align}\LABEQ{CROSS1}
&\E[\Delta R_{j,u}(\tau) ~ \Delta R_{z,x}(\tau)|\G(\tau)],\\\LABEQ{CROSS2}
&\E[\Delta V_u(\tau)~ \Delta V_x(\tau)|\G(\tau)],\\\LABEQ{CROSS3}
&\E[\Delta R_{j,u}(\tau)  ~  \Delta V_x(\tau)|\G(\tau)],
\end{align}
for $u,x\in[D]$ and $j,z\in[r]$.  Note that we do not compute second order moments taken across different time instants. As in Appendix\SEC{A1}, we omit the conditionality on $\G(\tau)$ and, to compute the expectations in \EQ{CROSS1}-\EQ{CROSS3}, we average over all possible positions where a degree-one check node can be removed. For instance,
\begin{align}
&\E[\Delta R_{j,u}(\tau) ~ \Delta R_{z,x}(\tau)]\\\nonumber
&=\sum_{m=1}^{D}\E[\Delta R_{j,u}(\tau) ~ \Delta R_{z,x}(\tau)|\mathtt{pos}(\tau)=m]p_m(\tau).
\end{align}
To evaluate $\E[\Delta R_{j,u}(\tau) ~ \Delta R_{z,x}(\tau)|\mathtt{pos}(\tau)=m]$, we consider two different scenarios: $u\neq x$ and $u=x$. 
%The first case is also split in two additional sub-cases: if $\mathtt{pos}(\tau)=m$, then we first consider the case $u\neq m$ and $x\neq m$. Second, we assume $\mathtt{pos}(\tau)=x$.
Namely, second order moments  across different positions and within the same position.

\vspace{5mm}
\subsection{Different positions. Case $u\neq x$} 

First, if $u\neq x$, then it is simple to check that
\begin{align}
\E[\Delta V_u(\tau)~ \Delta V_x(\tau)]=0,
\end{align}
since only one variable node is removed from the graph and it cannot belong to two distinct positions. Assume now that $x=u+c$, where $c$ is a strictly positive integer. If $c\geq l$, then the DD at positions $u$ and $x$ cannot be modified at the same time because check nodes at two positions further away $l-1$ positions do not share any variable node. Thus, for $c\geq l$
\begin{align}
\E[\Delta R_{j,u}(\tau) ~ \Delta R_{z,x}(\tau)|\mathtt{pos}(\tau)=m]&=0,\\
\E[\Delta R_{j,u}(\tau)  ~  \Delta V_x(\tau)|\mathtt{pos}(\tau)=m]&=0
\end{align}
for any $m\in[D]$.  For $c<l$, we consider the following two sub cases:

\subsubsection{$\mathtt{pos}(\tau)=m$, $u\neq m$, $x\neq m$}
\vspace{0.2cm}

%For the case $c<l$, %assume $\mathtt{pos}(\tau)=m$ and .
%
% As discussed in Section\SEC{A1},  in this case positions $u$ and $x$ are individually affected with probabilities $\xi_{m,u}(\tau)$ and $\xi_{m,x}(\tau)$, see \EQ{xi}. Nevertheless, note that if $c\geq l$, then both the DD at positions $u$ and $m$ cannot be modified at the same time. Check nodes at two positions further away of $l-1$ positions do not share any variable node and thus
%\begin{align}
%\E[\Delta R_{j,u}(\tau) ~ \Delta R_{z,u+c}(\tau)|\mathtt{pos}(\tau)=m]=0
%\end{align}
%for any $m\in\{1,\ldots,D\}$, $c\geq l$ and $\{j,z\}\in[1,r]$.
%%\begin{figure}[h]
%%\centering
%%\includegraphics[width=11cm]{Delta.pdf} 
%%\caption{Check nodes separated $\Delta$ positions share variable nodes placed at sections $[i+\Delta-\hat{l},\ldots,i+\hat{l}]$.}.
%%\LABFIG{Delta}
%%\end{figure}
%For $c<l$, 
The probability that the DD at positions $u$ and $x$ is simultaneously modified when we remove a degree-one check node from $m$ is
%we also require that the DD at positions $i$ and $x$ are both modified when we peel off the check node from $m$. This only happens as long as the removed variable node belongs to positions $[i+\Delta-(l-1),\ldots,i]$, which happens with probability
\begin{align}
\xi_{m,u,x}(\tau)=\sum_{i=x-(l-1)}^{u}\lambda_{m,i}(\tau).
\end{align}
and, therefore,
\begin{align}\nonumber
&\E[\Delta R_{j,u}(\tau) ~ \Delta R_{z,x}(\tau)|\mathtt{pos}(\tau)=m]=\\
\nonumber
&jz ~\xi_{m,u,x}(\tau)\left(\frac{R_{j+1,u}(\tau)-R_{j,u}(\tau)}{\sum_{q=1}^{r}R_{q,u}(\tau)}\right)\\
&\times  \left(\frac{R_{z+1,x}(\tau)-R_{z,x}(\tau)}{\sum_{q=1}^{r}R_{q,x}(\tau)}\right)\LABEQ{crossdelta}
\end{align}
for $j,z<r$. For $j=r$ ($z=r)$,  we delete from \EQ{crossdelta} the term in $j+1$ ($z+1$). Regarding the expected product $\Delta R_{j,u}(\tau)\Delta V_x(\tau)$, if $c<l$ then a variable node at position $x$ is connected to a check node at position $u$ with probability one. With probability $\lambda_{m,x}(\tau)$ the variable removed along with the degree-one check node at $m$ belongs to position $x$. Consequently, 
\begin{align}
&\E[\Delta R_{j,u}(\tau)\Delta V_x(\tau)|\mathtt{pos}(\tau)=m]\nonumber\\&=-j\lambda_{m,x}(\tau)\frac{R_{j+1,u}(\tau)-R_{j,u}(\tau)}{\sum_{q=1}^{r}R_{q,u}(\tau)}\LABEQ{crossRV1}.
\end{align}

\subsubsection{$\mathtt{pos}(\tau)=u$, $u\neq x$}
\vspace{0.2cm}

Since we remove a degree-one check node from position $u$, then $\Delta R_{j,u}(\tau)=0$ for any $j>1$:
\begin{align}
&\E[\Delta R_{j,u}(\tau) ~ \Delta R_{z,x}(\tau)|\mathtt{pos}(\tau)=ux]=0
\end{align}
for $z\in[r]$ and $j>1$. For $j=1$ we get
\begin{align}\LABEQ{eqz1}
&\E[\Delta R_{1,u}(\tau) ~ \Delta R_{z,x}(\tau)|\mathtt{pos}(\tau)=u]\nonumber\\
&=-z\xi_{u,x}(\tau)\frac{R_{z+1,x}(\tau)-R_{z,x}(\tau)}{\sum_{q=1}^{r}R_{q,x}(\tau)},
\end{align}
where for $z=r$ the term $R_{z+1,x}(\tau)$ in \EQ{eqz1} is zero. Similarly
\begin{align}\LABEQ{crossRV2}
&\E[\Delta R_{j,u}(\tau)\Delta V_x(\tau)|\mathtt{pos}(\tau)=u]=0
\end{align}
 for $j>1$ and 
 \begin{align}\LABEQ{crossRV3}
\E[\Delta R_{1,u}(\tau)\Delta V_x(\tau)|\mathtt{pos}(\tau)=u]=\lambda_{u,x}(\tau).
\end{align}

%
%Consider now the moments of the form
%\begin{align}\LABEQ{crooss3}
%\E[\Delta R_{j,u}(\tau)  ~  \Delta V_x(\tau)].
%\end{align}
%Clearly, since a variable at position $x=u+c$ is connected to check nodes at positions $x,\ldots,x+(l-1)$, \EQ{crooss3} is zero for $c\geq l$. For $c<l$, then the variable node at $x$ is connected to a check at position $u$ with probability one. Therefore,
%\begin{align}
%&\E[\Delta R_{j,u}(\tau)\Delta V_x(\tau)|\mathtt{pos}(\tau)=m]\nonumber\\&=-j\lambda_{m,x}(\tau)\frac{R_{j+1,u}(\tau)-R_{j,u}(\tau)}{\sum_{q=1}^{r}R_{q,u}(\tau)}\LABEQ{crossRV1}
%\end{align}
%for $u\neq m$. 

\vspace{5mm}
\subsection{Same positions. Case $x=u$}
Our goal now is to compute moments of the form
\begin{align}
\E[\Delta R_{j,u}(\tau) ~ \Delta R_{z,u}(\tau)]
\end{align}
for any position $u\in[D]$ and degrees $j,z\in[r]$. Assume with no loss of generality that $z\geq j$. The procedure is  similar to that followed in Appendix\SEC{A1}. First we assume that the position $m$ where we remove the degree-one check node is not position $u$, i.e., $\mathtt{pos}(\tau)=m\neq u$. After deleting the  variable node, one edge is removed from a check at position $u$ with probability $\xi_{m,u}(\tau)$. Besides, with probability 
\begin{align}
\displaystyle\frac{R_{j,u}(\tau)}{\sum_{q=1}^{r}R_{q,u}(\tau)}
\end{align}
the check node that was connected to such edge is of degree $j$. In this case,  $j$ edges of right degree $j$ are removed from the graph at position $u$ and $j-1$ edges of right degree $j-1$ are created at the same position. 
The rest of the DD terms are not affected. Therefore, 
\begin{align}\LABEQ{auxeq}
\E[\Delta R_{j,u}(\tau) ~ \Delta R_{z,u}(\tau)|\mathtt{pos}(\tau)=m]=0, ~~ z>j+1.
\end{align}
If $z=j+1$ we get
\begin{align}\LABEQ{auxeq2}
&\E[\Delta R_{j,u}(\tau) ~ \Delta R_{j+1,u}(\tau)|\mathtt{pos}(\tau)=m]\nonumber\\ &=-j (j+1)\xi_{m,u}(\tau)\frac{R_{j+1,u}(\tau)}{\sum_{u=1}^{r}R_{q,u}(\tau)},
\end{align}
and for $j=z$ we get
\begin{align}\LABEQ{auxeq3}
&\E[\Delta R_{j,u}(\tau)^2|\mathtt{pos}(\tau)=m]=j^2\xi_{m,u}(\tau)\frac{R_{j+1,u}(\tau)+R_{j,u}(\tau)}{\sum_{u=1}^{r}R_{q,u}(\tau)}.
\end{align}
For $j=r$, \EQ{auxeq2}  is also equal to zero while in \EQ{auxeq3}  the term $R_{j+1,u}(\tau)$ is set to zero. Finally, we compute
\begin{align}
&\E[\Delta^2 V_u(\tau)|\mathtt{pos}(\tau)=m]=\lambda_{m,u}(\tau),\\\nonumber\\
&\E[\Delta R_{j,u}(\tau)\Delta V_u(\tau)|\mathtt{pos}(\tau)=m]\nonumber\\&=-j\lambda_{m,u}(\tau)\frac{R_{j+1,u}(\tau)-R_{j,u}(\tau)}{\sum_{q=1}^{r}R_{q,u}(\tau)},\LABEQ{crossRV1}
\end{align}
where for the case $j=r$ we remove the term $R_{j+1,u}(\tau)$ from the equation above.

Now we assume $\mathtt{pos}(\tau)=u$. In this case, only one edge with right degree one is removed from position $u$. Then,
\begin{align}\LABEQ{auxeq}
\E[\Delta R_{j,u}(\tau) ~ \Delta R_{z,u}(\tau)|\mathtt{pos}(\tau)=u]=0
\end{align}
if $j>1$ or  $z>1$. For $z=j=1$:
\begin{align}
&\E[\Delta R_{1,u}(\tau)^2|\mathtt{pos}(\tau)=u]=1.
\end{align}

\section{Covariance Initial conditions}\LABSEC{A3}

In Section\SEC{PDEXPECTED}, we computed the expected DD of the
$(l,r,L)$ graph after the PD initialization step, $\E[V_{u}(0)]$
and $\E[R_{j,u}(0)]$ for $j\in[r]$ and $u\in[D]$. We are now
interested in computing the  initial conditions for the covariance
evolution equations described in Section\SEC{COVPD}, namely
$\text{Var}[R_{j,u}(0)]$, $\text{CoVar}[R_{j,u}(0),R_{z,x}(0)]$,
$\text{Var}[V_{u}(0)]$, $\text{CoVar}[V_{u}(0),V_{x}(0)]$ and
$\text{CoVar}[V_{u}(0),R_{j,x}(0)]$ for $(j,z)\in[r]^2$ and $(u,x)\in[D]^2$.
For simplicity, we will consider positions $u,x$ that belong both to the
range $[l,\ldots,L]$, where all check nodes in the original $(l,r,L)$
graph are of degree $r$. The rest of moments, involving positions
at the boundaries, can be computing in a similar procedure.

After PD initialization,  the number of check nodes of degree $1, 2,\ldots,r$ at position $u\in[l,L]$ is described by a multinomial distribution with $\frac{l}{r}M$ trials and probabilities
\begin{align}\LABEQ{pju}
p_{j,u}=\binom{r}{j}\pe^j(1-\pe)^{r-j}
\end{align}
for $j\in[r]$, see Section\SEC{PDEXPECTED}. Therefore, 
\begin{align}
\text{Var}[R_{j,u}(0)]&=j^2\frac{l}{r}Mp_{j,u}(1-p_{j,u}),\\
\text{CoVar}[R_{j,u}(0),R_{z,u}(0)]&=-jz\frac{l}{r}Mp_{j,u}p_{z,u}.
\end{align}
In addition, since variable nodes are independently erased
\begin{align}
\text{Var}[V_{u}(0)]&=M\epsilon (1-\epsilon),\\
\text{CoVar}[V_{u}(0),V_{x}(0)]&=0\quad u\neq x.
\end{align} 
Now we focus on the computation of moments of the form
\begin{align}\LABEQ{crossR}
\text{CoVar}[R_{j,u}(0),R_{z,x}(0)]
\end{align}
for $x\neq u$. First, if positions $|u-x|\geq l$ then by construction
of the $(l,r,L)$ ensemble a variable node cannot be connected
simultaneously to one check node at position $u$ and to one check
node at position $x$. Consequently, $\text{CoVar}[R_{j,u}(0),R_{z,x}(0)]=0$
for any pair of degrees $(j,z)$ as long as $|u-x|\geq l$. For a similar reason,
if $|u-x|\geq l$ then $\text{CoVar}[V_{u}(0),R_{j,x}(0)] =0$.

The main idea to evaluate \EQ{crossR} when $|u-x|< l$ is to compute
the probability that any pair of check nodes (one at position $u$
and one at position $x$) selected at random share at least one
variable node in the $(l,r,L)$ code graph, i.e., before PD initialization. In such a case, the corresponding degrees of both
check nodes after the PD initialization are not statistically
independent from each other. Consider  $x=u+c$ where $c<l$. There
are $(l-c)$ positions of the code, from position $x-l+1$ to position
$u$, in which any variable node is connected with one edge to a
check node at position $u$ and with one edge to a check node at
position $x$. Let $\mathtt{check}_u$ and $\mathtt{check}_x$ be a
pair of check nodes selected at random from positions $u$ and $x$
and let $a$ ($b$) represent the number of edges of the check 
$\mathtt{check}_u$ ($\mathtt{check}_x$) that are connected  to variable nodes at positions in the range
$\{x-l+1,\ldots,u\}$.

Recall from the properties of the $(l,r,L)$ ensemble described in
Section\SEC{S1} that, if we picked at random one edge connected to
a check node placed at position $u$, the position of the variable
node connected to such edge is a uniform random variable in the set
$\{u-(l-1), \ldots,u\}$.  Hence, among the $r$ edges connected to
$\mathtt{check}_u$, the number $a$ of edges that are connected to
variables in the range of positions $\{x-l+1,\ldots,u\}$ is  a
random variable distributed according to a Binomial distribution
of $r$ trials and success probability $(l-c)/l$. The same holds for
$b$. Also, $a$ and $b$ are independent random variables.

For any given pair $(a,b)$, the probability that $\mathtt{check}_u$ and $\mathtt{check}_x$ are connected to the same variable node, placed at a position in the range $\{x-l+1,\ldots,u\}$, is as follows:
\begin{align}\LABEQ{probconect}
(l-c)\frac{ab}{M}.
\end{align}
By averaging \EQ{probconect} over all possible pairs $(a,b)$ we can compute the probability  that $\mathtt{check}_u$ and $\mathtt{check}_x$ share one variable node:
\begin{align}
P_S=\frac{l-c}{M}r^2\left(\frac{l-c}{l}\right)^2.
\end{align}
In the following, we ignore the case that $\mathtt{check}_u$ and $\mathtt{check}_x$ share two variable nodes since the probability of this to happen decays by $M^{-2}$.  Denote by $d_{u}$ and $d_{x}$ the degree of the check nodes  $\mathtt{check}_u$ and $\mathtt{check}_x$ respectively after PD initialization.  The joint probability mass function of the  pair $(d_u,d_x)$ can be expressed as follows:
\begin{align}\LABEQ{jointdegrees}
P(d_u=j,d_x=z)&=P(d_u=j,d_x=z|\text{share})P_S\\
&+P(d_u=j,d_x=z|\text{no share})(1-P_S)\nonumber,
\end{align}
where $P(d_u=j,d_x=z|\text{share})$ represents the joint probability distribution of the degrees $d_u$ and $d_x$ when $\mathtt{check}_u$ and $\mathtt{check}_x$ share one variable node. It is straightforward to show that
\begin{align}
&P(d_u=j,d_x=z|\text{share})\\
&=\epsilon\left[\binom{r-1}{j-1}\epsilon^{j-1}(1-\pe)^{r-j}\binom{r-1}{z-1}\epsilon^{z-1}(1-\pe)^{r-z}\right]\nonumber\\
&+(1-\epsilon)\left[\binom{r-1}{j}\epsilon^{j}(1-\pe)^{r-j-1}\binom{r-1}{z}\epsilon^{z}(1-\pe)^{r-z-1}\right]\nonumber,
\end{align}
and 
\begin{align}
&P(d_u=j,d_x=z|\text{no share})=p_{j,u}p_{z,x},
\end{align}
where $p_{j,u}$ and $p_{z,x}$ are given in \EQ{pju}. Using these expressions and averaging over all possible $(\mathtt{check}_u,\mathtt{check}_x)$ pairs (there are $(\frac{l}{r} M)^2$ pairs in total), we obtain:
\begin{align}\LABEQ{Cov}
&\text{CoVar}[R_{j,u}(0),R_{z,x=u+c}(0)]=jz M (l-c)^3 \\\nonumber
&\times \Big(P(d_u=j,d_x=z|\text{share})-P(d_u=j,d_x=z|\text{no share})\Big)
\end{align}
for $u,x\in[l,\ldots,L]$, $j,z\in[r]$ and $c<l$. By following a similar procedure, we obtain
 \begin{align}\LABEQ{Cov2}
&\text{CoVar}[V_{u}(0),R_{j,x=u+c}(0)] \nonumber \\
&=j M \pe^j(1-\pe)^{r-j}\left[\binom{r-1}{j-1}-\binom{r}{j}\pe\right]
\end{align}
for $u,x\in[l,\ldots,L]$ and $c<l$. 

%For $c\geq l$, the covariance in \EQ{Cov2} is zero since the DDs at positions $u$ and  $x=u+c$ are independent. In this appendix, we have restricted the computations of covariance moments of the DD $\G(0)$ involving positions in the $(l,r,L)$ ensemble where the check degree is constant to $r$. The covariance moments relating positions  $[1,\ldots,l-1]$ or $[L+1,\ldots,L+l-1]$ can be computed in a similar way.

%\begin{figure*}[!t]
%% ensure that we have normalsize text
%\normalsize
%\begin{align}\LABEQ{jointdegrees}
%P(d_u=j,d_x=z)=P(d_u=j,d_x=z|\text{share})P_S+P(d_u=j,d_x=z|\text{no share})(1-Pv)
%\end{align}
%% Restore the current equation number.
%%\setcounter{equation}{\value{mytempeqncnt}}
%% IEEE uses as a separator
%\hrulefill
%% The spacer can be tweaked to stop underfull vboxes.
%\vspace*{4pt}
%\end{figure*}

\end{document}